\documentclass[aps,prl,twocolumn,reprint,showpacs,groupedaddress]{revtex4-1}
\usepackage{amsmath,amssymb,graphicx,color}
\usepackage{times}
\usepackage[table]{xcolor}
\usepackage{amsmath}
\usepackage{graphicx}
\usepackage{amssymb}
\usepackage{txfonts,color,cancel}
\usepackage[colorlinks=true,linkcolor=blue]{hyperref}

\newcommand\ket[1]{\left|\textstyle{#1}\right\rangle}
\newcommand\bra[1]{\left\langle\textstyle{#1}\right|}
\newcommand\braket[1]{\left\langle\textstyle{#1}\right\rangle}
\newcommand\half{\frac{1}{2}}
\newcommand\down{\downarrow}

\newcommand\sig{\sigma}
\newcommand\adag{a^\dagger}

\newcommand\adaga{a^\dagger a}

\newcommand\Hz{\textrm{Hz}}
\newcommand\kHz{\textrm{kHz}}
\newcommand\MHz{\textrm{MHz}}

\begin{document}
\title{Probing the Dynamics of Superradiant Quantum Phase Transition in a Single Trapped-Ion}
\author{Ricardo Puebla, Myung-Joong Hwang, Jorge Casanova, and Martin B. Plenio}
\affiliation{Institut f\"{u}r Theoretische Physik and IQST, Albert-Einstein-Allee 11, Universit\"{a}t Ulm, D-89069 Ulm, Germany}
\begin{abstract}
We demonstrate that the quantum phase transition (QPT) of the Rabi model and critical dynamics
near the QPT can be probed in the setup of a single trapped ion. We first show that there exists
equilibrium and non-equilibrium universal functions of the Rabi model by finding a proper rescaling
of the system parameters and observables. We then propose a scheme that can faithfully realize the
Rabi model in the limit of a large ratio of the effective atomic transition frequency to the oscillator
frequency using a single trapped-ion and therefore the QPT. It is demonstrated that the predicted
universal functions can indeed be observed based on our scheme. Finally, the effects of realistic noise
sources on probing the universal functions in experiments are examined.
\end{abstract}

\maketitle
{\it Introduction.---} The experimental realization of quantum phase transition (QPT) in a
well-controlled quantum system is of considerable interest~\cite{Greiner:2002es,Bloch:2012jy,Porras:2004di,Kim:2010ib,Islam:2011ct,Bermudez:2012hs,Baumann:2010js,Dimer:2007da,Nagy:2010dr}.
This is particularly important for the study of the dynamics of QPT where a controlled change of
the system parameters are necessary~\cite{Damski:2005cr,Zurek:2005cu,Polkovnikov:2005gr}. Understanding
the dynamics of QPT is at the frontier of the study of critical phenomena; the full extent of the
universality in non-equilibrium dynamics of a system that undergoes a QPT remains to be determined~\cite{Braun:2015iw,Klinder:2015df}
and its theoretical underpinnings are being actively investigated~\cite{Polkovnikov:2011iu,Eisert:2015ka,Kolodrubetz:2012jo,Nikoghosyan:2016ff,Acevedo:2014eo,Hwang:2015eq}.

Trapped ions are a particularly promising platform for this purpose thanks to the possibility of
precise coherent quantum controls and high-fidelity measurements~\cite{Porras:2004di,Kim:2010ib,Islam:2011ct}
as exemplified by the recent observation of the dynamics of classical phase transitions
\cite{pyka2013topological,ulm2013observation}. A major challenge, however, lies in the fact
that the QPT typically occurs in a thermodynamic limit where the number of system constituents
diverges~\cite{Sachdev:2011uj}. Although the universality manifests itself even for a system of
finite size in the form of finite-size scaling relations~\cite{Fisher:1972tv,Botet:1982ju}, it
emerges only when the system size is sufficiently large; moreover, a controlled change in the
system size under otherwise unchanged conditions is necessary in order to observe the
critical exponents. In this respect, and despite the advances in trapped-ion technologies, it
is still a formidable challenge to scale up the system size sufficiently to enable the observation
of critical phenomena while maintaining the controllability and the coherence of the system~\cite{Islam:2011ct}.

Recently, it has been shown in Ref.~\cite{Hwang:2015eq, Hwang:2016vf} that even a single two-level
atom coupled to a harmonic oscillator may undergo a second-order QPT. The experimental realization 
of such a finite-system QPT is highly desirable, as it opens a possibility to study the dynamics of 
QPT in a small, fully controlled quantum system with a high degree of coherence without the necessity 
of the scalability in the number of system components; however, the required parameter regime~\cite{Hwang:2015eq, Hwang:2016vf} 
that includes simultaneously extremely large detuning~\cite{Ashhab:2013ke,Bakemeier:2012ja,Hwang:2010jn}
and large coupling strength~\cite{Bourassa:2009gy, Ashhab:2010eh,Casanova:2010kd} has made it difficult 
to find a suitable experimental platform to realize the finite-system QPT.

In this letter, we demonstrate that the QPT of the Rabi model, as well as universal dynamics of this QPT,
can be observed experimentally with a {\it single} trapped-ion. Before discussing a trapped-ion realization,
we first demonstrate the existence of a non-equilibrium universal function for the adiabatic dynamics
of the Rabi QPT which goes beyond a power-law behavior predicted by Kibble-Zurek mechanism~\cite{Hwang:2015eq,Damski:2005cr,Zurek:2005cu,Polkovnikov:2005gr}.
Interestingly, we show that while the latter is difficult to observe directly in the trapped ion setup,
the former is readily accessible under the same conditions. Moreover, for the equilibrium QPT, we find
a scaling function for the atomic population of the ground state, which we propose to use to measure the
finite-size scaling exponent of the ground state.

%that we propose to use as a probe of universal dynamics of Rabi QPT, as well as a theory of equilibrium scaling function for the probe of ground state scaling relations. The universal functions that we find here take the power-law behaviors in adiabatic dynamics predicted by the Kibble-Zurek mechanism and the finite-frequency critical exponents shown in Ref.~\cite{Hwang:2015eq} as a limiting case. Interestingly, we find that the the exponents for Kibble-Zurek dynamics and the critical exponents for the ground state are difficult to observe directly in the trapped ion setup, while it is feasible to observe the universal functions. Therefore, not only is our study of the universal functions important in the fundamental point of view for understanding the universal physics of the Rabi model in and out of equilibrium, but they are also proved to be a crucial tool for experimentally probing the QPT in the trapped-ion setup.

We then consider a concrete trapped-ion realization where the Rabi model is realized
by dichromatic sideband lasers such that the atom-coupling strength can be modulated
by the intensity of the lasers while the atomic and oscillator frequency can be chosen
by the frequency of the lasers~\cite{Cirac:1993db,Pedernales:2015cj}. However, we show that,
in the limit of our interest where the critical behavior emerges, the standard approach based
on traveling-wave lasers~\cite{Cirac:1993db,Pedernales:2015cj} cannot faithfully realize the
Rabi model and obscures its universal behavior. We propose and analyse a standing-wave
configuration~\cite{Cirac:1992ck,deLaubenfels:2015fb} taking into account current experimental
limitations and show that it can overcome current limitations to achieve a high-fidelity
realization of the Rabi model. More specifically, by solving the dynamics of the single
trapped-ion in a standing wave configuration where the Rabi frequencies associated with the
applied lasers is adiabatically changed, we demonstrate that it is indeed possible to observe
the universal functions, predicted in the first part of the letter, in a realistic trapped-ion setup.

Finally, we examine the effect of different noise sources in our proposed ion trap
realisation on probing the universal functions. It is shown that the non-equilibrium
universal function is noise-resilient thanks to the short adiabatic evolution time
that is required. The equilibrium scaling function, however, in general turns out
to be strongly affected by the noise and thus difficult to observe; nevertheless, we show
that its asymptotic behavior is still unaffected by the noises, which allows us to
measure quantitatively the finite-frequency scaling exponent.

{\it Finite-frequency scaling.---} The Rabi Hamiltonian reads
\begin{equation}
    \label{eq:01}
    H_\textrm{Rabi}=\omega_0\adag  a+\frac{\Omega}{2}\sigma_z-\lambda( a+\adag) \sigma_x
\end{equation}
where $\sigma_{x,z}$ are the Pauli matrices for a two-level atom and $a$ ($a^\dagger$) is
an annihilation (creation) operator for a cavity field. The oscillator frequency is $\omega_0$,
the atomic transition frequency $\Omega$, and the coupling strength $\lambda$. We introduce a
dimensionless coupling constant $g=2\lambda/\sqrt{\omega_0\Omega}$ and the frequency ratio
$R=\Omega/\omega_0$. In the $R\rightarrow\infty$ limit, the Rabi model undergoes a second-order
QPT at the critical point $g=1$~\cite{Hwang:2015eq}. For large but not infinite $R$,
the ground state expectation values and the energy spectrum exhibit a critical scaling in
$R$, so-called {\it finite-frequency} scaling~\cite{Hwang:2015eq}, near the critical point. Here we focus
on the ground state population of the two-level atom, $\braket{\sigma_z}$, because it is
possible to measure it with a high-fidelity in the trapped-ion system~\cite{Myerson:2008bm,Burrell:2010kh},
and we derive the analytic expression for its scaling relations. Below we discuss the main
results, while we refer to the Supplemental Material~\cite{sup} for the detailed derivation.

\begin{figure}[t]
\centering
\includegraphics[width=\linewidth,angle=0]{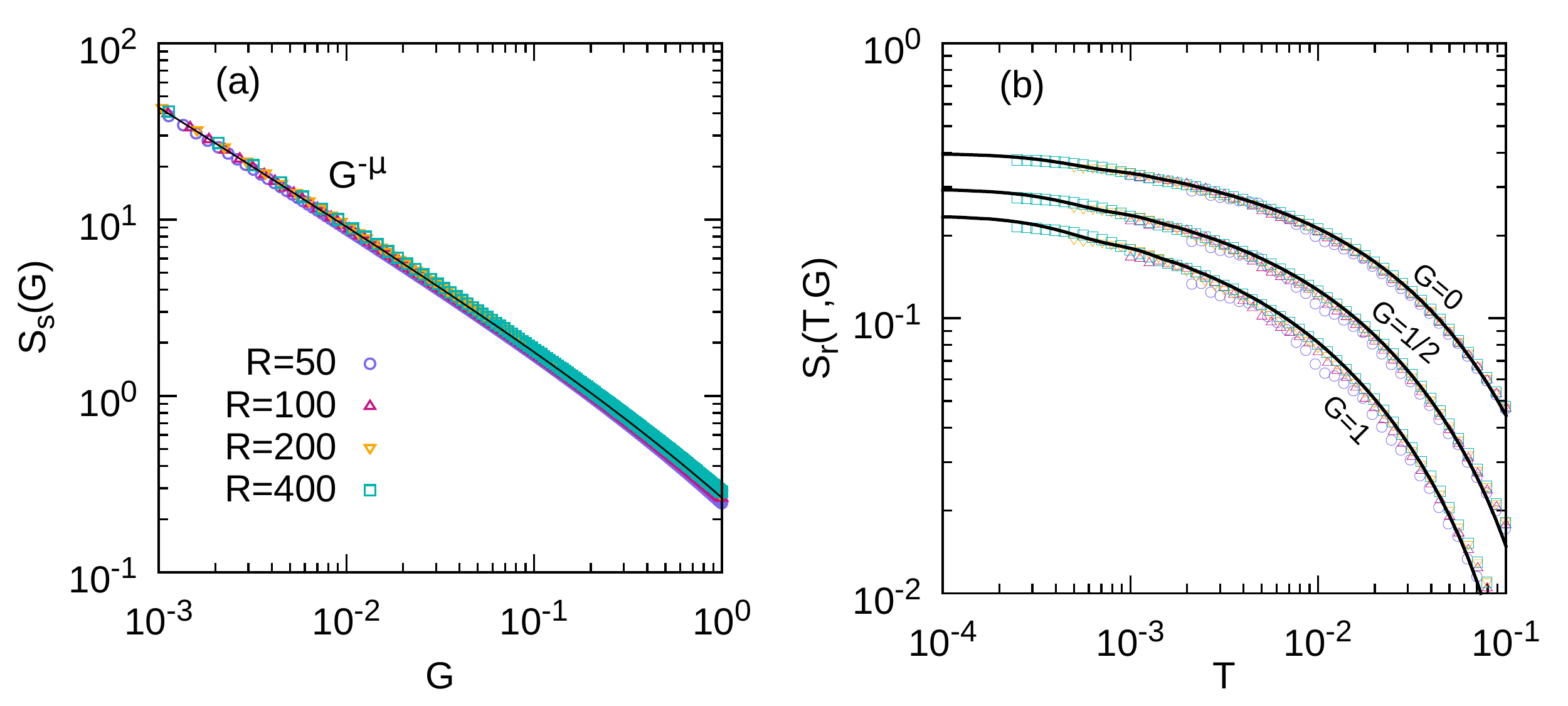}
\caption{Universal functions of the Rabi model. (a) The rescaled ground state population
$S_s$ as a function of the rescaled coupling strength $G=R|g-1|^{\gamma/\mu}$ with $\gamma=1$
and $\mu=2/3$ for different frequency ratio $R=50$, $100$, $200$, and $400$ and coupling
strength $0.9\leq g\leq1$. For $G\ll1$, it follows a power-law $S_s\propto G^{-\mu}$. (b)
The rescaled residual atomic population $S_r$ as a function of a rescaled quench time
$T=R^{-1}\tau_q$ for a fixed value of the rescaled final coupling strength
$G_f=R|g_f-1|^{\gamma/\mu}=0$, $1/2$, and $1$. The quench time $\tau_q$ varies from
$0.1/\omega_0$ to $100/\omega_0$ and $0.9\leq g_f \leq1$. The same values of $R$ as in (a) has been used.}
\label{fig:1}
\end{figure}

In the $R\rightarrow\infty$ limit we have $\braket{\sigma_z}=-1$ for $g\leq1$ and $\braket{\sigma_z}
=-\frac{1}{g^2}$ for $g>1$~\cite{Hwang:2015eq}. Then, the singular part of $\braket{\sigma_z}$
is $\braket{\sigma_z}_\textrm{s}\equiv\braket{\sigma_z}+1=(1-g^{-2})$ and it vanishes as
$\braket{\sigma_z}_\textrm{s}\propto(g-1)^\gamma$ near the critical point  with a critical
exponent $\gamma=1$. We now consider the singular part of $\braket{\sigma_z}$ as a function
of $R$ and $g$, denoted by $\braket{\sigma_z}_s(R,g)$, and examine specifically its scaling behavior
for finite $R$. Particularly, we find the analytical expression of the finite-frequency scaling
for $R\gg1$ and $g=1$ as
\begin{align}
    \label{eq:02}
    \braket{\sigma_z}_s(R,g=1)&\propto R^{-\mu}
\end{align}
where $\mu=2/3$ is the finite-frequency scaling exponent of $\sigma_z$. See the Supplemental
Material~\cite{sup} for the derivation of Eq.~(\ref{eq:02}) based on a variational method, as 
well as its excellent agreement with the numerical results. Furthermore, by a rescaling of the 
expectation value and coupling strength as
\begin{align}
    \label{eq:03}
    S_s\equiv |g-1|^{-\gamma}\braket{\sigma_z}_s,\quad G\equiv R|g-1|^{\gamma/\mu},
\end{align}
we find that the ground state population of the spin can be cast into a universal
form, $S_s(G)$, which is called as a scaling function~\cite{Fisher:1972tv,Botet:1982ju}.
The functional form can be obtained by (i) calculating $\braket{\sigma_z}_s(R,g)$
using the numerically exact diagonalization for different values of $R$ and $g$ satisfying
$R\gg1$ and $|g-1|\ll1$ and (ii) plotting the rescaled quantity $S_s=|g-1|^{-\gamma}\braket{\sigma_z}_s$
as a function of $G=R|g-1|^{\gamma/\mu}$. As shown in Fig.~\ref{fig:1} (a), all the data
points collapse onto a single curve, confirming the existence and revealing the functional
form of $S_s(G)$.  We also find an analytic expression for the asymptotic behavior of
$S_s(G)$ as $\lim_{G\rightarrow0} S_s(G)\propto G^{-\mu}$, which agrees very well with
the numerical results [Fig.~\ref{fig:1} (a)]. We emphasize that the asymptotic behavior
of $S_s(G)$ for $G\ll1$ is governed by the finite-frequency scaling exponent $\mu$ shown
in Eq.~(\ref{eq:02}).

{\it Adiabatic evolution and dynamical scaling.---} We now consider an adiabatic dynamics of
the Rabi model. To this end we prepare an initial state $\ket{\Psi(t=0)}=\ket{0}\ket{\down}$,
where $\ket{0}$ is the zero phonon Fock state and $\ket{\down}$ is an atomic eigenstate, and
increase the coupling strength linearly in time from $g_i=0$ to $g_f$ for a duration $\tau_q$,
that is, $g(t)=g_ft/\tau_q$. By setting $\tau_q$ to be large enough to satisfy the adiabatic
condition~\cite{Messiah:1961vw}, one can prepare the ground state of the Rabi model with
$g=g_f$ for a fixed $R$ to high fidelity and measure $\sigma_z$ to observe the ground state
universal function $S_s(G)$ and the scaling exponent $\mu$ discussed in the previous section.
A potential limit to this approach is that the spectral gap $\Delta$ vanishes at the QPT
($R\rightarrow \infty$) as $\Delta\propto\left|g-1\right|^{\zeta}$ where $\zeta=1/2$ is its
critical exponent~\cite{Sachdev:2011uj,Hwang:2015eq} and, for a finite $R$, the gap $\Delta$
of $H_\textrm{Rabi}$ decays as a power-law $\Delta\propto R^{-\mu\zeta/\gamma}=R^{-1/3}$ near
the critical point; therefore, for a large value of $R$, the adiabatic condition will
require $\tau_q$ to be much larger than the coherence time of the system. While we examine
the feasibility of the adiabatic preparation in the last section in much more detail, here we
consider the case when $\tau_q$ becomes progressively smaller; then, the adiabatic condition
starts to break down near the critical point first, while it is still satisfied away from the
critical point. In other words, we go beyond the equilibrium setting and examine the universality
in adiabatic dynamics of the QPT.

The key insight for the adiabatic dynamics of the QPT is that due to the critical scaling of
the equilibrium properties, e.g., shown in Eq.~(\ref{eq:02}), one can cast the equation of
motion for the adiabatic evolution into a universal form through a rescaling of parameters
\cite{Kolodrubetz:2012jo,Acevedo:2014eo,Nikoghosyan:2016ff}. For the Rabi model, we find that
by rescaling the evolution time $\tau_q$ as
\begin{equation}
    T\equiv R^{-\frac{\gamma}{\mu(1+\zeta)}}\tau_q,
\end{equation}
where $\gamma/\mu(1+\zeta)=1$, and together with the rescaling the coupling strength
$G=R|g-1|^{\gamma/\mu}$ already introduced in Eq.~(\ref{eq:03}), the equation of motion
transforms into a universal form~\cite{sup}, which does not depend on the specific values
of system parameters $R$ or $g$. As in the case of the ground state QPT, the central quantity
of our interest is the population of the two-level system. Let us denote $\braket{\sigma_z}_f(R,g_f,\tau_q)$
as the expectation value of $\sigma_z$ for the final state of the adiabatic evolution for
a given quench time $\tau_q$ at the final coupling strength $g_f$ and the frequency ratio
$R$, and $\braket{\sigma_z}(R,g_f)$ as the ground state expectation value of $\sigma_z$ for
a given $R$ and $g=g_f$. Now, we introduce a quantity that we call the residual atomic population
as $\braket{\sigma_z}_r(R,g_f,\tau_q)\equiv|\braket{\sigma_z}_f(R,g_f,\tau_q)-\braket{\sigma_z}(R,g_f)|$,
which quantifies the non-adiabaticity of the evolution and vanishes for a $\tau_q$
satisfying the adiabatic condition. Our main result is that the rescaled residual atomic
population $S_r\equiv R^{\mu}\braket{\sigma_z}_r$ is a universal function of the rescaled
parameters $T$ and $G$~\cite{sup}. To confirm this prediction, we solve the dynamics for
different $\tau_q$ and calculate the residual atomic population for a set of values of
$g_f$ and $R$ leading to a fixed value of $G_f= R|g_f-1|^{\gamma/\mu}$. Then, we plot the
rescaled residual atomic population $S_r$ as a function of T and show that all the data
points with the same value of $G_f$ collapse into a single curve confirming that $S_r(T,G_f)$
is a universal function [Fig.~\ref{fig:1} (b)]. It is clear that different choices of $G_f$
lead to different universal curves, as $S_r$ is a function of both $G_f$ and $T$ [Fig.~\ref{fig:1} (b)].

{\it Trapped-ion realization.---}
Having established both the equilibrium and non-equilibrium scaling theory of the Rabi
model in terms of the atomic population, here we demonstrate that it is feasible to probe
the predicted scaling functions [Fig.~\ref{fig:1}] in a trapped-ion setup. We consider a
setup of a single trapped ion with two traveling-wave lasers, described by the following
Hamiltonian,
\begin{equation}
    \label{TI}
    H_\textrm{TI}(t)=\nu a^\dagger a+\frac{\omega_I}{2}\sigma_z+\sum_{j=1,2}\frac{\Omega^d_j}{2}\sigma_+e^{i(\eta_j(a+a^\dagger)-\omega^d_jt-\phi^d_j)}+\textrm{H.c.},
\end{equation}
where $a~(\sigma_-)$ is an annihilation operator for a phonon (internal levels). The phonon
frequency is $\nu$ and the transition frequency $\omega_I$. For $j$-th laser, $\omega^d_j$ is
the frequency, $\phi^d_j$ the phase, $\Omega^d_j$ the Rabi frequency, and $\eta_j=k_j\sqrt{1/(2m\nu)}$
is the Lamb-Dicke parameter with $k_j$ being the wave vector and $m$ the ion mass. We consider
two driving frequencies to be tuned near the blue-sideband and red-sideband transition,
respectively, i.e., $\omega_{1,2}^d=\omega_I\mp\nu+\delta_{1,2}$, where $\delta_{1,2}\ll\nu$
are additional detunings w.r.t each sidebands and we set $\Omega^d_1=\Omega^d_2=\Omega^d$,
$\eta_1=\eta_2=\eta$ and $\phi_1^d=\phi_2^d=3\pi/2$. Note that the so-called optical rotating-wave
approximation has already been made to Eq.~(\ref{TI}), which is well-known to hold in this setting.

In the rotating frame with respect to $H_\textrm{TI}^0=(\nu-\tilde\omega_0) a^\dagger a +
\half(\omega_I-\tilde\Omega)\sigma_z$, Eq.~(\ref{TI}) becomes time-independent and assumes
the form of the Rabi model~\cite{Cirac:1993db,Pedernales:2015cj},
\begin{equation}
    \label{TI_rf}
    \tilde H_\textrm{TI}(t)=e^{-iH_\textrm{TI}^0t} H_\textrm{TI}(t)e^{iH_\textrm{TI}^0t}\simeq\tilde\omega_0\adag  a+\frac{\tilde\Omega}{2}\sigma_z-\tilde\lambda( a+\adag) \sigma_x.
\end{equation}
Here the new set of parameters for the Rabi model are
\begin{equation}
    \label{newrabi}
    \tilde\omega_0=\half(\delta_1-\delta_2),\quad\tilde\Omega=\half(\delta_1+\delta_2),\quad\tilde\lambda=\eta\Omega^d/2,
\end{equation}
which are determined by the frequencies and the amplitudes of the applied lasers. We emphasize
that Eq. (\ref{TI_rf}) is valid only within the Lamb-Dicke regime, i.e., $\eta\sqrt{\langle(a+a^\dagger)^2\rangle}\ll1$
and the vibrational rotating-wave approximation (RWA), therefore it is not \textit{a priori}
evident that one can probe the QPT of the Rabi model using this approach. The in and out
of equilibrium universal properties of the Rabi model emerge when $R=\tilde \Omega/\tilde \omega_0\gg1$
and $g=2\tilde\lambda/\sqrt{\tilde\omega_0\tilde\Omega}\simeq1$. However, the phonon population in
the ground state of the Rabi model monotonically increases as one increases $\tilde\Omega/\tilde\omega_0$,
leading to a potential departure from the Lamb-Dicke regime. Furthermore, the strong coupling strength
$\tilde\lambda\simeq\sqrt{\tilde\omega_0\tilde\Omega}$ requires a large Rabi frequency of the laser, which
could break the vibrational rotating wave approximation. Hence we now need to study in detail whether
the desired regime $R\gg 1$ and $g\cong 1$ can indeed be reached.

\begin{figure}[t]
\centering
\includegraphics[width=\linewidth,angle=0]{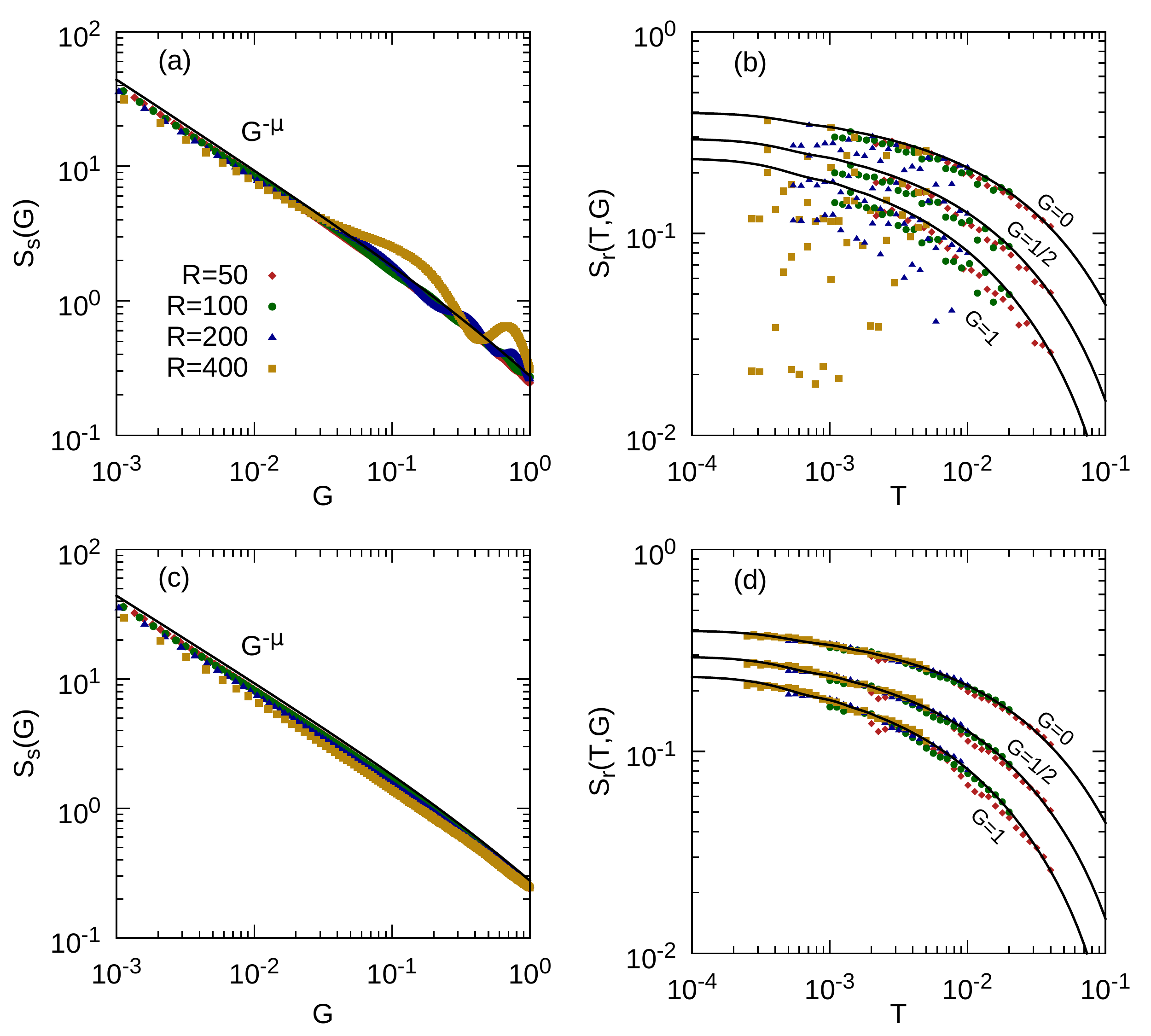}
\caption{Universal functions $S_s(G)$ and $S_r(T,G)$ obtained from the trapped-ion Hamiltonians with
the scheme of (a), (b) two lasers (c), (d) six lasers. The different points corresponds to for different
frequency ratios, $R=50$, $100$, $200$ and $400$ with a fixed effective oscillator frequency
$\tilde{\omega}_0/2\pi=200\textrm{Hz}$. The solid lines is the result of the Rabi model presented in
Fig.~\ref{fig:1}. In (a) and (c), the quench time is chosen as $\tau_q=50/\tilde{\omega}_0=250\textrm{ms}$
to adiabatically prepare the ground state and obtain $S_s(G)$; meanwhile, in (b) and (d), a range of
quench time $0.5\textrm{ms}\leq\tau_q\leq10\textrm{ms}$ for three different values $G=0$, $1/2$ and $1$
is used to obtain the dynamical scaling function $S_r(T,G)$ }
\label{fig:2}
\end{figure}

In order to examine whether the Rabi QPT can be probed in the trapped-ion setup, we apply the adiabatic
protocol discussed in the previous section directly to the trapped-ion Hamiltonian in Eq.~(\ref{TI}) without
assuming any simplification, neither the Lamb-Dicke nor the vibrational RWA. This involves preparing the
initial state $\ket{\Psi(t=0)}=\ket{0}\ket{\down}$ where $\ket{0}$ is the zero-phonon Fock state and
$\ket{\down}$ is the low-energy state of the ion, and adiabatically turning on the Rabi frequency
$\Omega^d(t)$ of the two lasers for a duration of $\tau_q$ until it reaches the desired final value
of $g=g_f$; that is, $\Omega^d(t)=\Omega^d_f t/\tau_q$ with $\Omega^d_f=g_f\sqrt{\delta_1^2-\delta_2^2}/2\eta$,
while the detunings $\delta_{1,2}$ are chosen to realize a fixed value of $R$ and remain fixed during the
adiabatic evolution. Then, one measures the population of the internal levels $\sigma_z$ of the final states
of the adiabatic evolution, from which one finds the universal scaling function $S_s(G)$ and $S_r(T,G)$
through the rescaling of the parameters as described in the previous section.

A possible set of parameters for the Rabi model realized in the trapped ion set up would be
$\tilde\omega_0/2\pi=200\Hz$ and $10\kHz\leq\tilde\Omega/2\pi\leq 80\kHz$ so that the frequency
ratios  $50\leq R \leq 400$ can be explored. This in turn implies that the Rabi frequency at the
critical point $g=1$ is $23.6 \kHz\leq \Omega^d/2\pi \leq 66.6 \kHz$, respectively, where we have
used the Lamb-Dicke parameter $\eta=0.06$. For the adiabatic preparation of the ground state, the
considered evolution time is $\tau_q= 50/\tilde{\omega}_0=250 \textrm{ms}$ (below we discuss
possible drawbacks concerning long-time evolutions) which approximately satisfies the adiabatic
condition~\cite{sup}. Meanwhile, for the dynamical scaling, one can choose a smaller range,
$0.1/\tilde{\omega}_0\leq\tau_q\leq2/\tilde{\omega}_0$, or equivalently, $0.5\textrm{ms}\leq\tau_q\leq 10\textrm{ms}$.
The numerical results with the above parameters are shown in Fig.~\ref{fig:2} (a) and (b). We observe
a strong deviation from the theoretical prediction of the Rabi model and the rescaled expectation values
do not collapse into the predicted universal function.

We identify that a leading order contribution to the breakdown of the Eq.~(\ref{TI_rf}) is a carrier
interaction, i.e., $-i\frac{\Omega^d}{2}(\sigma_+e^{i\delta_jt}-\sigma_-e^{-i\delta_jt}$) for $j=1,2$,
that is induced by the both sideband transitions due to the large Rabi frequency $\Omega^d$ used to
achieve a strong coupling strength $\tilde\lambda$. The effect of this spurious process becomes dominant
for $R\gg1$ and obscures the universality of the Rabi model [Fig.~\ref{fig:2} (a,b)]. To resolve
this issue, we propose to use a standing wave configuration for the sideband lasers so that the carrier
interaction is suppressed in the leading order. That is, we consider two additional lasers in the
Eq.~(\ref{TI}), labeled as $j=3,4$, such that $\omega_3^d=\omega_1^d$ and $\omega_4^d=\omega_2^d$
and $\eta_1=-\eta_3$ and $\eta_2=-\eta_4$. The phases are chosen as $\phi_3=\phi_4=\pi/2$, so that
the ion is located at node of the standing wave when the Rabi frequencies of two counter-propagating
lasers are set to be identical $\Omega_{3,4}^d=\Omega_{1,2}^d=\Omega^d$.

Note that in the standing wave configuration, the effective coupling strength becomes
$\tilde \lambda=\eta\Omega^d$. This means that to reach the critical point $g=1$, one
would need $11.8 \kHz\leq \Omega^d/2\pi \leq 33.3 \kHz$ which is a factor $2$ smaller
than the traveling wave configuration. With the same parameters used for the
travelling-wave configuration except for the Rabi frequencies, we present the
numerical results of the adiabatic evolution with the standing-wave configuration
described above in [Fig.~\ref{fig:2} (c) and (d)]. Here, we have taken into account
small, but experimentally inevitable differences in the Rabi frequencies of the
counter-propagating lasers, i.e., $\Omega_{3,4}\neq\Omega_{1,2}$. Even with $8\%$
of error in the Rabi frequencies leading to an imperfect standing wave, the results
show an excellent agreement with the prediction of the Rabi model [Fig.~\ref{fig:2}
(c) and (d)] and demonstrate that it is possible to probe the universal scaling
functions of the Rabi model in a trapped-ion setup with the standing wave configuration.

\begin{figure}[t]
\centering
\includegraphics[width=\linewidth,angle=0]{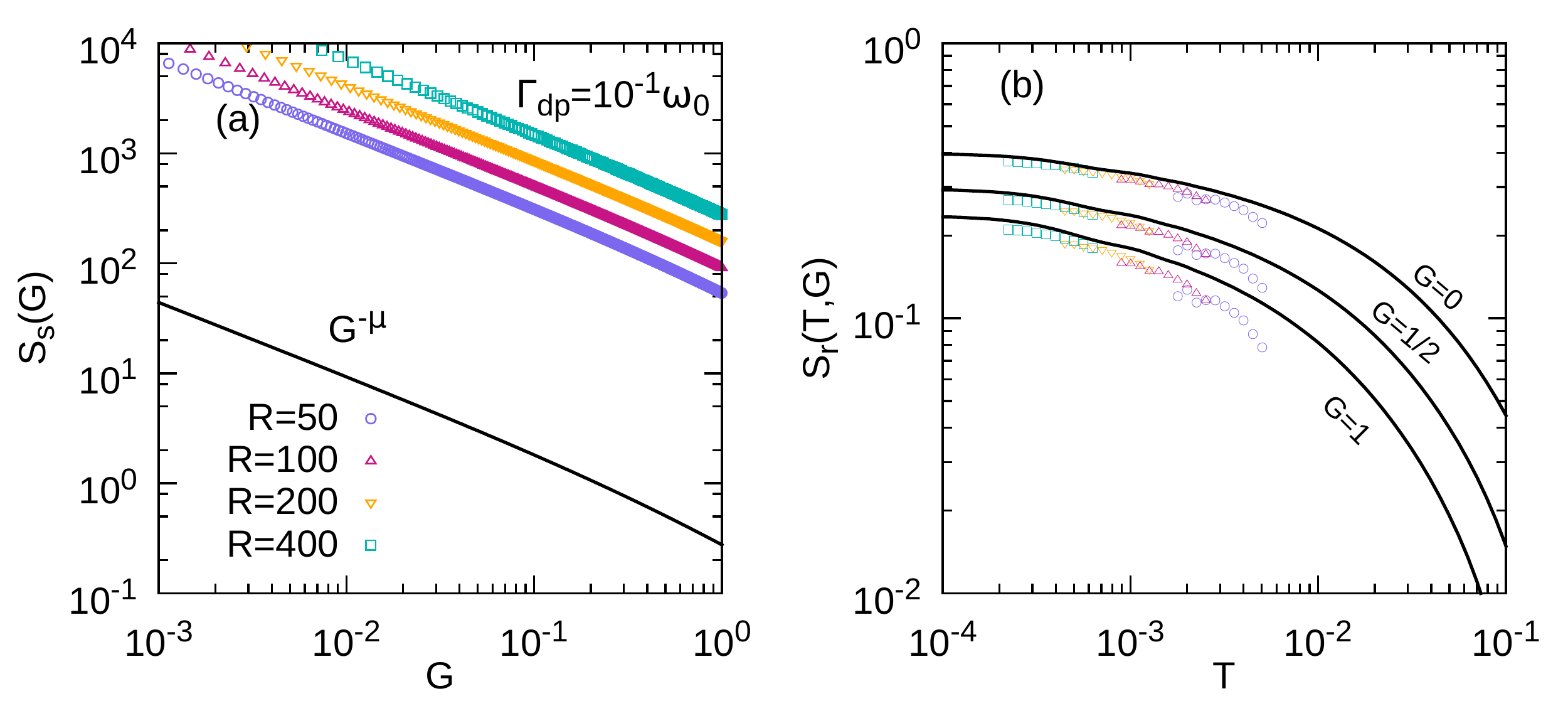}
\caption{Effect of noises on the universal functions. The universal functions are obtained
from the solution of master equation with a dephasing rate of $\Gamma_{dp}=0.1\omega_0$ and
decay and heating rates of $\Gamma_{c}=\Gamma_a=\Gamma_h=0.05\omega_0$  (see the main text
for further details). The same system parameter used in the Fig.~\ref{fig:1} is considered
and the solid lines correspond to the Rabi model results as in Fig.~\ref{fig:1}.  (a) While
the graphs corresponding to the different system size $R$ do not collapse, the asymptotic
scaling $G^{-\mu}$ for $G\ll1$ is preserved.  (b) The dynamical scaling function is robust
against noise. Note that we have constrained the range of $\tau_q$ to rather short times,
$0.1/\omega_0\leq\tau_q\leq 0.275/\omega_0$ as the longer time evolution deviates from the
universal behavior due to the effect of noise~\cite{sup}; while this leads to smaller data
points than Fig.~\ref{fig:1} and \ref{fig:2}, it nevertheless correctly reveals the substantial
part of the universal function. }
\label{fig:3}
\end{figure}

{\it Effects of noise.---} During the adiabatic evolution, various experimental noise sources
will have an impact on the final states. Here we examine the effects of noise on the
dynamics and demonstrate that the predicted universal functions can be observed in the
realistic experimental conditions. The master equation that governs the adiabatic evolution is
\begin{align}
    \dot \rho=&-i[H_\textrm{Rabi}(t),\rho]+\Gamma_{dp}\mathcal{L}[\sigma_z] +
    \Gamma_c\mathcal{L}[\sigma_-]\nonumber \\&+\Gamma_a\mathcal{L}[a]+\Gamma_h\mathcal{L}[a^{\dagger}],
\end{align}
where $\mathcal{L}[x]=x\rho x^{\dagger}-x^{\dagger}x\rho/2-\rho x^{\dagger}x/2$ is the
Lindbladian superoperator. A typical parameter for dephasing of the internal states of
the ion is $\Gamma_\textrm{dp}/2\pi=20\Hz$ and the rest can be estimated as
$\Gamma_{c}/2\pi=\Gamma_{a}/2\pi=\Gamma_h/2\pi=10\textrm{Hz}$. Therefore, we set
$\Gamma_\textrm{dp}/\omega_0=0.1$ and $\Gamma_\textrm{c}/\omega_0=\Gamma_\textrm{a}/\omega_0
=\Gamma_\textrm{h}/\omega_0=0.05$ and solve the adiabatic dynamics with the same parameter
sets used in the previous sections.

As shown in Fig.~\ref{fig:3} (a), the ground state scaling function $S_s(G)$ is in fact
strongly influenced by the effect of noise. For the different frequency ratio $R$, the graphs
no longer collapse onto the theoretically predicted universal function. This comes from the
fact that the long evolution time $\tau_q\sim 250\textrm{ms}$ required for the adiabatic
condition is much larger than the coherence time of the ion, which is about $50\textrm{ms}$.
Interestingly, however, the asymptotic behavior of $S_s(G)$ for $G\ll1$ still follows the
predicted power-law of $G^{-\mu}$. Therefore, our simulation shows that one can quantitatively
measure the finite-frequency scaling exponent of $\sigma_z$ of the ground state, even in the
presence of the noises. The measurement of the finite-frequency scaling exponent could serve
as an experimental confirmation of the quantum phase transition of the Rabi model~\cite{Hwang:2015eq}.

On the other hand, we find that the non-equilibrium universal function $S_r(T,G)$ is much
more robust to the effect of noise. As shown in Fig.~\ref{fig:3} (b), the rescaled data
points collapse into a single universal curve. Although there is a slight deviation from
the ideal case, the universality in the dynamics still remains intact. The robustness of
the non-equilibrium universal function to noises stems from the relatively short evolution
time $\tau_q$ compared to the coherence time. While the small spectral gap near the critical
point necessitates a very large $\tau_q$ for the adiabatic preparation of the ground state,
the nearly adiabatic dynamics considered here requires the adiabaticity only away from the
critical point where the energy gap does not vanish, which makes its experimental observation
more favorable than the equilibrium case.

{\it Conclusion.---} We have demonstrated that the Rabi QPT and its universal dynamics can
be observed in a trapped-ion setup with a single ion using the equilibrium and non-equilibrium
universal functions as a probe thus opening the doors for the experimental exploration of
the properties of second-order quantum phase transitions with trapped ions.

{\it Acknowledgements.---} This work was supported by an Alexander von Humboldt Professorship,
the ERC Synergy grant BioQ, the EU STREP project EQUAM and the CRC TRR21. J.C. acknowledges support to the Alexander von Humboldt foundation

\pagebreak
\widetext
\begin{center}
\textbf{ \large Supplemental Material: \\Probing the Dynamics of Superradiant Quantum Phase Transition in a Single Trapped-Ion}
\end{center}
\setcounter{equation}{0}
\setcounter{figure}{0}
\setcounter{table}{0}

%\makeatletter
\renewcommand{\theequation}{S\arabic{equation}}
\renewcommand{\thefigure}{S\arabic{figure}}
\renewcommand{\bibnumfmt}[1]{[S#1]}
\renewcommand{\citenumfont}[1]{S#1}

\section{Section A: Ground-state finite-frequency scaling analysis for $\braket{\sigma_z}$ }
\label{sec:A}
\begin{figure}[t]
\centering
\includegraphics[width=0.5\linewidth,angle=0]{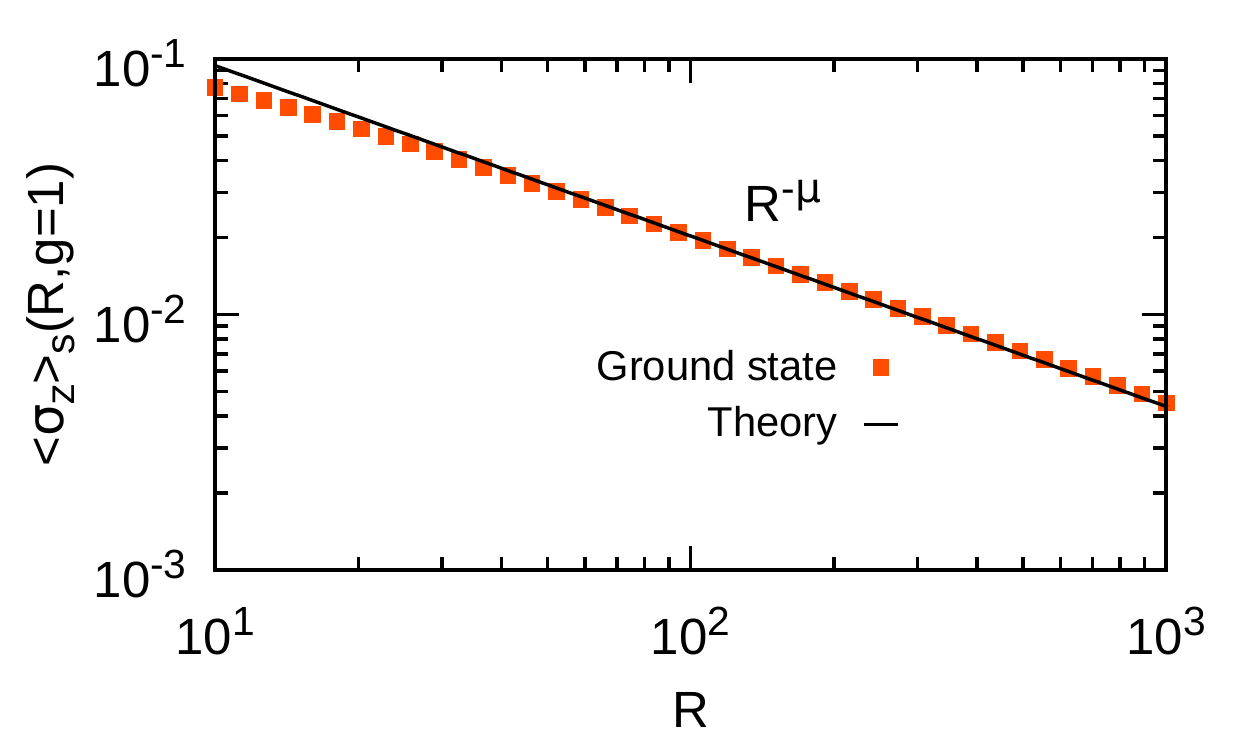}
\caption{Scaling of $\braket{\sigma_z}_s(R,g=1)$ at the critical point as a function of the frequency ratio $R$. The orange squares correspond to numerical diagonalization of $H_{\textrm{Rabi}}$ while the solid line to the analytical expression. The power-law scaling perfectly agrees with the theoretical prediction $\mu=2/3$ already for $R\gtrsim 50$.}
\label{fig:szScaling}
\end{figure}

Here we present a derivation for the finite-frequency scaling relation for the ground state of the Rabi Hamiltonian, particularly the atomic population $\braket{\sigma_z}$, shown in Eq.~(2) of the main text. Following the Ref.~\cite{Hwang:2015eqs}, we first apply a unitary transformation to the Rabi Hamiltonian which decouples the spin subspace up to the fourth order in the coupling strength $\lambda$, that is,
\begin{align}
H'=U^\dagger H_\textrm{Rabi} U=\omega_0a^\dagger a +\frac{\Omega}{2}\sigma_z+\frac{\omega_0g^2}{4}(a+a^{\dagger})^2\sig_z-\frac{g^4\omega_0^2}{16\Omega}(a+a^\dag)^4\sig_z+\mathcal{O}\left(\frac{g^2\omega_0^3}{\Omega^2}\right),
\end{align}
where 
\begin{equation}
U=\exp\left[\frac{\lambda}{\Omega}(a+a^\dagger)\left(1-\frac{4\lambda^2}{3\Omega^2}(a+a^\dagger)^2\right)(\sigma_+-\sigma_-)\right].
\end{equation}
An approximate solution for the ground state wave function of $H'$ at $g=1$ can be obtained by a variational method~\cite{Hwang:2015eqs}, which leads to
\begin{equation}
\ket{\phi'_G(R,g=1)}=\mathcal{S}[s(R)]\ket{0}\ket{\down}
\end{equation}
where $\mathcal{S}[s]=\exp[{\frac{s}{2}(a^{\dagger 2}-a^2)}]$ and 
\begin{equation}
s(R)=\frac{1}{6}\ln\left(\frac{2}{3}R\right).
\end{equation}
From this variational solution, we obtain the expectation value of $\sigma_z$ for the ground state of the Rabi Hamiltonian as
\begin{align}
\label{eq:s1}
\braket{\sigma_z}(R,g=1)&=\bra{\down}\bra{0}\mathcal{S}^\dagger[s(R)]U^\dagger \sigma_z U\mathcal{S}[s(R)]\ket{\down}\ket{0}\nonumber\\
&=\bra{\down}\bra{0}\mathcal{S}^\dagger[s]\left(\sigma_z(1-\frac{1}{2R}(a+a^\dagger)^2)+\sigma_x\frac{1}{\sqrt{R}}(a+a^\dagger)+\mathcal{O}((\omega_0/\Omega)^{3/2})\right)\mathcal{S}[s]\ket{\down}\ket{0}\nonumber\\
&\simeq-1+\frac{1}{3}\left(\frac{2}{3}R\right)^{-2/3}.
\end{align}
It follows that
\begin{align}
 \braket{\sigma_z}_s(R,g=1)=\braket{\sigma_z}(\Omega/\omega_0,g=1)+1\propto R^{-2/3},
\end{align}
which is the Eq.~(2) of the main text. In the Fig.~\ref{fig:szScaling}, we confirm the above prediction using the numerically exact diagonalization of the Rabi Hamiltonian. For the frequency ratio $R>50$, $\braket{\sigma_z}(R,g=1)$ does show a power-law with exponent $\mu=2/3$.

\section{Section B: non-equilibrium scaling analysis for $\braket{\sigma_z}$ }
\label{sec:B}
\begin{figure}[t]
\centering
\includegraphics[width=0.75\linewidth,angle=0]{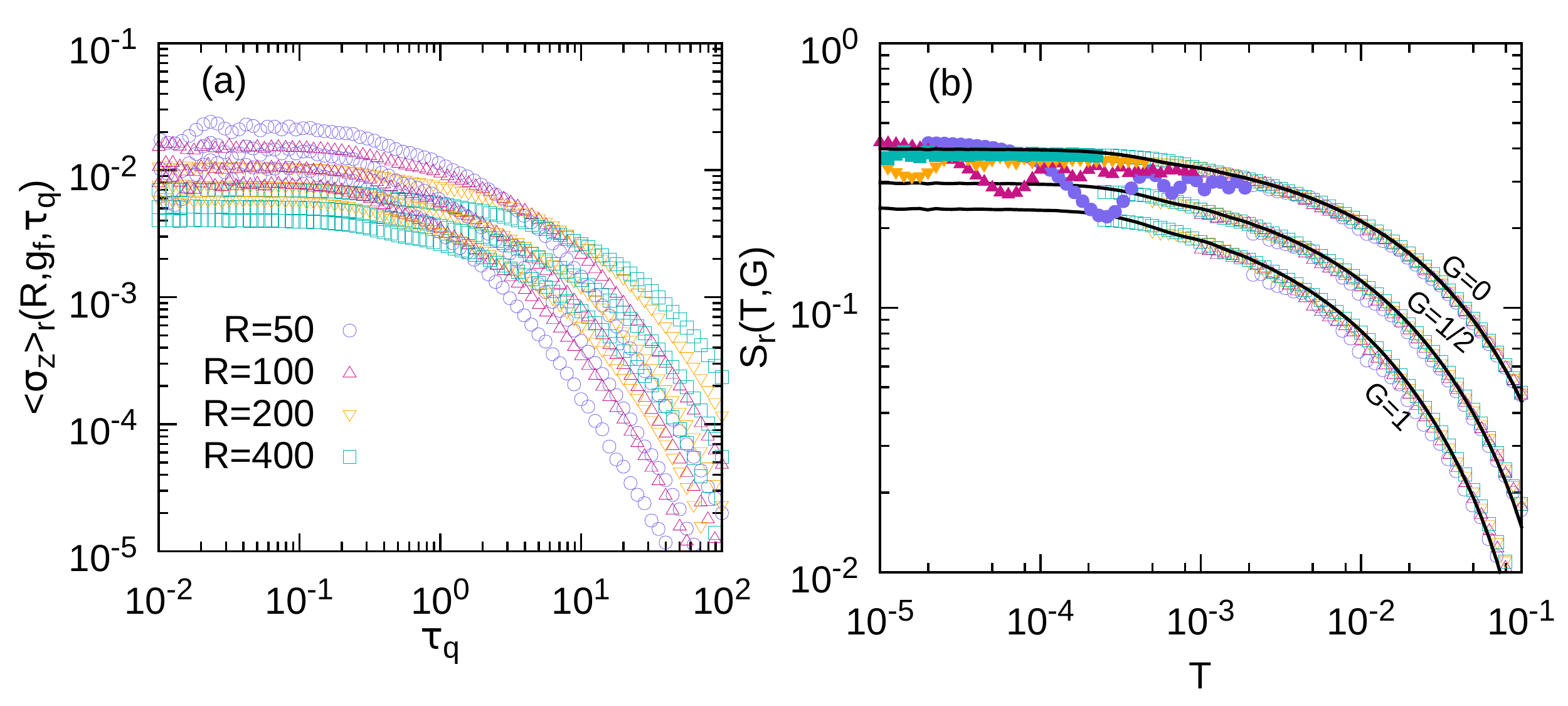}
\caption{In (a) the residual atomic population $\braket{\sigma_z}_r(R,g_f,\tau_q)$ for $R=50$, $100$, $200$ and $400$ as a function of the quench time $\tau_q$ and $g_f$ corresponding to  $G=0$, $1/2$ and $1$. In (b) the same results are depicted with the rescaling of the parameters to show the dynamical scaling function $S_r(T,G)$ as a function of $T$. For too fast quenches, $\omega_0\tau_q\ll1$, non-universal oscillations emerge since excitations are not mainly created close to the QPT. This is illustrated for $G=0$ with filled points ($\tau_q<0.1/\omega_0$).}
\label{fig:NonEqszScaling}
\end{figure}

% Brief der. of non-eq scaling
% motivate the way of rescaling
% Fig: <\sigma_z> [bare dynamics] --> collapse [include results for \tau_q<0.1/w0 (sudden dynamics)]
In this section, we derive a proper rescaling of parameters that casts an equation of motion for the adiabatic evolution into a universal form in the $R\gg1$ limit following the procedure of Ref.~\cite{Acevedo:2014eos}. This rescaling is used to reveal the non-equilibrium universal function for the residual population of the ion introduced in the main text. Recall that we consider a linear driving $g(t)=g_ft/\tau_q$, in which the rate $\dot{g}=g_f/\tau_q$ is controlled by the quench time $\tau_q$. The wave function at time $t$ can be written as
\begin{align}
\ket{\Psi(t)}=\sum_{n=0}^\infty c_n(t)e^{-i\int_{t_0}^tE_n(t')dt'}\ket{\phi_{g(t)}^n},
\end{align}
where $\ket{\phi_{g(t)}^n}$ are the instantaneous eigenstates of the Rabi Hamiltonian at $g(t)$, i.e., $H_{\textrm{Rabi}}(g(t))\ket{\phi_{g(t)}^n}=E_n(g(t))\ket{\phi_{g(t)}^n}$. The Schr{\"o}dinger equation can then be written in terms of  $c_n(g)$, \begin{align}
\frac{d}{dg}c_n(g)=\sum_{m\neq n}e^{-i\frac{\tau_q}{g_f}\int_{g_0}^{g}\Delta_{n,m}^R(g')dg'}\chi^R_{n,m}(g)c_m(g),
\end{align}
where $\Delta_{n,m}^R(g)$ and $\chi_{n,m}^R(g)$ are the energy difference and transition amplitude between the $n$th and $m$th eigenstates for a given frequency ratio $R$, respectively. The latter is given by $\chi_{n,m}^R(g)=-\left<\phi_g^n \right| \partial_g\left|\phi_g^m \right>$. Both quantities $\Delta_{n,m}^R(g)$ and $\chi_{n,m}^R(g)$ follow a finite-frequency scaling relation for $|g-1|\ll1$ and $R\gg1$,
\begin{align}
%\Delta_{n,m}^R(g)=R^{-\mu\zeta/\gamma}F_{\Delta_{n,m}}(G)\\
%\chi_{n,m}^R(g)=R^{\gamma/\mu}F_{\chi_{n,m}}(G)
\Delta_{n,m}^R(g)=\left|g-1\right|^{\zeta}F_{\Delta_{n,m}}(G),\\
\chi_{n,m}^R(g)=\left|g-1\right|^{\beta}F_{\chi_{n,m}}(G),
\end{align}
where $G\equiv R\left|g-1\right|^{\gamma/\mu}$, and $F(G)$ is their corresponding finite-frequency scaling function. Note that the critical exponents are given by $\gamma=1$, $\mu=2/3$, $\zeta=1/2$ and $\beta=-1$~\cite{Hwang:2015eqs}. Assumming that $\tau_q$ is sufficiently large so that the main non-adiabatic excitations are formed close to the critical point and using the above finite-frequency scaling relation, the equation of motion can be rewritten as
\begin{align}
\frac{d}{dG}c_n(G)=\sum_{m\neq n}e^{-i \frac{\tau_q}{g_f}R^{-\mu(1+\zeta)/\gamma}\frac{\mu}{\gamma}\int_{G_0}^{G_f}F_{\Delta_{n,m}}(G')dG'}F_{\chi_{n,m}}(G)c_m(G)\frac{\mu}{\gamma G}.
\end{align}
Now, it is immediate that by rescaling the quench time $\tau_q$ as $T\equiv R^{-\gamma/(\mu(1+\zeta))}\tau_q=R^{-1}\tau_q$, the equation of motion can be made  to depend only on two rescaled variables, $G$ and $T$. It is universal in a sense that the equation of motion, and thus the solution for $c_n$, do not depend on the specific values of $R$, $g_f$ or $\tau_q$. Using this result, we can construct dynamical scaling functions for different observables. We focus on the atomic population $\braket{\sigma_z}$. The residual atomic population is defined as
\begin{align}
\label{res}
\braket{\sigma_z}_r(R,g_f,\tau_q)&\equiv\left|\braket{\sigma_z}_f(R,g_f,\tau_q)-\braket{\sigma_z}(R,g_f)\right|,
\end{align}
where $\braket{\sigma_z}_f(R,g_f,\tau_q)=\left<\Psi(\tau_q) \left|\sigma_z \right|\Psi(\tau_q) \right>$ is the final population at the end of quench and $\braket{\sigma_z}(R,g_f)=\left<\phi_{g_f}^0 \right|\sigma_z \left|\phi_{g_f}^0\right>$ is the ground-state expectation value for a given frequency ratio $R$ and final coupling $g_f$. Note that Eq.~(\ref{res}) can be also written in terms of $\braket{\sigma_z}_s=\braket{\sigma_z}+1$. Then, we have
\begin{align}
\braket{\sigma_z}_r(R,g_f,\tau_q)&=\left| \sum_{n=0}^\infty\left|c_n(\tau_q)\right|^2 \left<\phi_{g_f}^n \right|\sigma_z \left|\phi_{g_f}^n\right>_s -\left<\phi_{g_f}^0 \right|\sigma_z \left|\phi_{g_f}^0\right>_s \right|\nonumber\\
&\approx \left|\left<\phi_{g_f}^0 \right|\sigma_z \left|\phi_{g_f}^0\right>_s\right| S_r(c_0(\tau_q),c_1(\tau_q),....)\nonumber\\
&\approx R^{-\mu}S_r (T,G)
\end{align}
where we have used the fact that $\left<\phi_{g_f}^n \right|\sigma_z \left|\phi_{g_f}^n\right>_s\sim(2n+1)\left<\phi_{g_f}^0 \right|\sigma_z \left|\phi_{g_f}^0\right>_s$ for $g_f\sim g_c$. Therefore, the dynamical scaling function for $\braket{\sigma_z}_r(R,g_f,\tau_q)$ is obtained as $S_r(T,G)=R^{\mu}\braket{\sigma_z}_r(R,g_f,\tau_q)$ where $T\equiv R^{-\gamma/(\mu(1+\zeta))}\tau_q$ and $G\equiv R\left|g-1\right|^{\gamma/\mu}$.  In the panel (a) of Fig.~\ref{fig:NonEqszScaling}, we show the bare dynamics of $\braket{\sigma_z}_r(R,g_f,\tau_q)$ for $R=50$, $100$, $200$ and $400$ as a function of the quench time $\tau_q$ and for three different values of $g_f$ which correspond to $G=0$, $1/2$ and $1$. There is no collapse in the bare dynamics, as expected. In the panel (b) of Fig.~\ref{fig:NonEqszScaling}, however, by rescaling the parameters as prescribed above, there emerges universal curves onto which all the different curves collapses. When the  quench time is too short, $\omega_0\tau_q\ll 1$, the collapse does not occur even after rescaling. This is because the analysis for the adiabatic evolution is no longer valid. The open points correspond to $\tau_q>0.1/\omega_0$ and the full points at $G=0$ correspond to $\tau_q<0.1/\omega_0$. Therefore, $S_r(T,G)$ is obtained for quenches with $\tau_q>0.1/\omega_0$.
%\sum_{n=0}\left|c_n(t)\right|^2\left<\phi_{g_f}^n \right|\sigma_z \left|\phi_{g_f}^n\right>$  and $\braket{\sigma_z}(R,g_f)=\left<\phi_{g_f}^0 \right|\sigma_z \left|\phi_{g_f}^0\right>$

\section{Section C: Condition for adiabatic evolution}
\label{sec:C}
\begin{figure}[t]
\centering
\includegraphics[width=0.75\linewidth,angle=0]{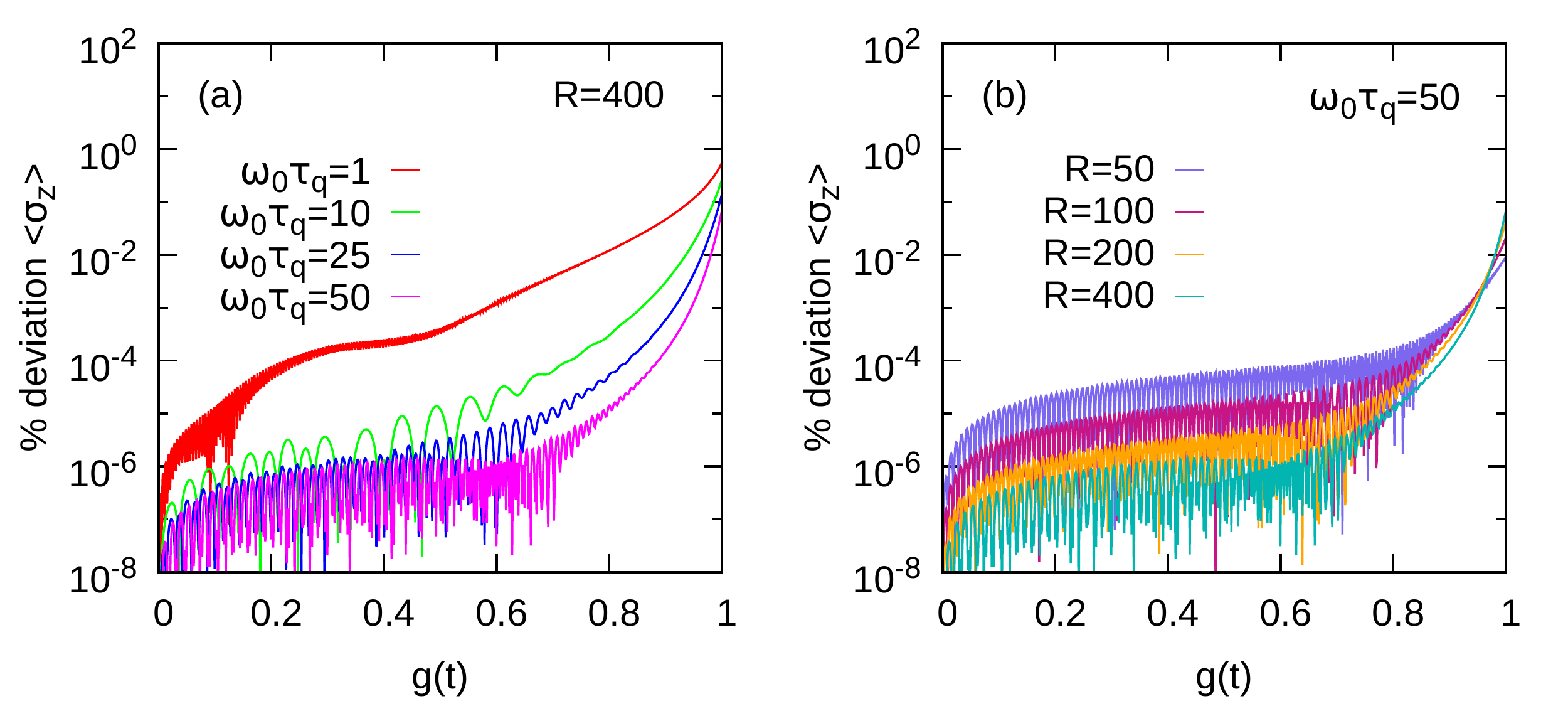}
\caption{Relative difference, in $\%$, among $\braket{\sigma_z}(t)$ and the ground state $\braket{\sigma_z}_{GS}$ as a function of the coupling parameter $g(t)$. In (a) the relative difference for $R=400$ for different quench rates, $\omega_0\tau_q=1$, $10$, $25$ and $50$. In (b) the relative difference for different values of $R$ and fixed quench time $\omega_0\tau_q=50$. As $g(t)$ approaches to the critical point $g=1$, the deviation increases, but it remains always below $0.1\%$ at any $g(t)$.}
\label{fig:figAdsz}
\end{figure}

Here we determine the optimal quench time $\tau_q$ for the adiabatic preparation of the ground state at the critical point. The state is initialized as $\ket{\Psi(0)}=\ket{0}\ket{\downarrow}$, which is the ground state of $H_{\textrm{Rabi}}$ at $g=0$. We then solve the unitary dynamics and calculate $\braket{\sigma_z}(t)=\left< \Psi(t) \right|\sigma_z\left|\Psi(t) \right>$ for different $R$ and quench times $\tau_q$, and then compare it with the expectation value of the instantaneous ground state $\braket{\sigma_z}_{GS}$. We choose the quench time $\tau_q$ for which the relative difference among $\braket{\sigma_z}(t)$ and $\braket{\sigma_z}_{GS}$ is lower than a certain tolerance value. However, since the energy gap vanishes at the critical point in the  $R\rightarrow\infty$ limit~\cite{Hwang:2015eqs,Sachdev:2011ujs}, the larger $R$, the longer the quench time $\tau_q$  will be needed to keep the deviation below a particular tolerance. We plot for $R=400$ the deviation for different quench times $\tau_q$ in Fig.~\ref{fig:figAdsz} (a). As one can observe, the adiabatic preparation gets worse as $g(t)$ approaches to the critical point. However, for $\tau_q= 50/\omega_0$, the relative difference drops below $0.1\%$, which is already reasonable good. Therefore, we set $\tau_q=50/\omega_0$ in order to achieve an adiabatic preparation of the ground state within $0.1\%$ tolerance error. In Fig.~\ref{fig:figAdsz} (b), for a fixed $\tau_q=50/\omega_0$, the relative difference is shown as a function of $R$, where $R=400$ is the worst case at $g=1$.

\section{Section D: Trapped-Ion realization of Rabi model}
\label{sec:D}
%derivation of Rabi model (mention spurious carrier)
%Errors in standing wave configuration
In this section we present the derivation of the Rabi model realization in a trapped-ion setting, based first on traveling waves~\cite{Cirac:1993dbs,Pedernales:2015cjs} and then, we extend the scheme to a standing-wave configuration~\cite{Cirac:1992cks,deLaubenfels:2015fbs}. A trapped ion of two internal levels, separated by $\omega_I$, is confined in a harmonic trap of frequency $\nu$, and the laser beams induce a coupling between them. The Hamiltonian can be written as $(\hbar=1)$
\begin{align}
\label{eq:hti}
H_{\textrm{TI}}(t)=\nu \adaga+\frac{\omega_I}{2}\sigma_z+\sum_j\frac{\Omega_j^d}{2}\sigma_x\left[ e^{i \left(k_j\hat{x}-\omega_j^d t -\phi_j^d\right)}+\textrm{H.c.}\right],
\end{align}
where the $j$th laser is characterized by its Rabi frequency $\Omega_j^d$ , frequency $\omega_j^d$, wave vector $k_j$ and phase $\phi_j^d$. The ion position operator is $\hat{x}=x_0(a+\adag)$, with $x_0=\sqrt{1/(2m\nu)}$ where $m$ is the ion mass. The Lamb-Dicke parameter is defined as $\eta_j=k_j x_0$.
In a rotating frame with respect to $H_0=\nu \adaga+\frac{\omega_I}{2}\sigma_z$, the previous Hamiltonian reads
\begin{align}
H^I=e^{iH_0t}H_{\textrm{TI}}(t)e^{-iH_0t}=\sum_j\frac{\Omega_j^d}{2}\left(\sigma_+e^{i\omega_It}+\textrm{H.c}\right)\left(e^{i\left(\eta_j(ae^{-i\nu t}+\adag e^{i\nu t})-\omega_j^d t-\phi_j^d \right)}+\textrm{H.c}\right).
\end{align}
Since we are interested in a parameter regime where $\omega_I-\omega_j^d\approx \pm \nu$, while $\omega_I+\omega_j^d\gg 1$, and also $\Omega_j^d\ll \omega_I+\omega_j^d$, we can safely perform an optical rotating-wave approximation (RWA), which neglects the terms that rotate at frequency $\omega_I+\omega_j^d$. This leads to an approximate Hamiltonian
\begin{align}
H^I\approx\sum_j\frac{\Omega_j^d}{2}\left(\sigma_+e^{i\left(\eta_j(ae^{-i\nu t}+\adag e^{i\nu t})+(\omega_I-\omega_j^d) t-\phi_j^d \right)}+\textrm{H.c}\right).
\end{align}
Assuming the Lamb-Dicke regime, $\eta\sqrt{\langle \left(a+\adag \right)^2\rangle}\ll 1$, we have
\begin{align}
e^{i\eta(ae^{-i\nu t}+\adag e^{i\nu t})}=I+i\eta\left(ae^{-i\nu t}+\adag e^{i\nu t}\right)+\mathcal{O}\left(\eta^2\right).
\end{align}
To realize the Rabi model in the trapped-ion setting, we need to introduce two traveling waves with frequencies $\omega_{1,2}^d=\omega_I\pm \nu -\delta_{1,2}$, where $\delta_{1,2}\ll \nu$, giving rise to a blue and red sideband interaction, respectively. This leads to
\begin{align}
H^I\approx &\frac{\Omega_1^d}{2}\left(\sigma_+\left[I+i\eta_1\left(ae^{-i\nu t}+\adag e^{i\nu t}\right) \right]e^{i((-\nu+\delta_1)t-\phi_1^d)} +\textrm{H.c.}\right)+\nonumber\\
+&\frac{\Omega_2^d}{2}\left(\sigma_+\left[I+i\eta_2\left(ae^{-i\nu t}+\adag e^{i\nu t}\right)\right]e^{i((\nu+\delta_2)t-\phi_2^d)} +\textrm{H.c.}\right).
\end{align}
At this point, we invoke the RWA to neglect the terms that rotate at frequency $\nu\sim \MHz$, which is valid for a relatively small Rabi frequency $\Omega_j^d\sim \kHz$. This approximation is called vibrational RWA, after which we obtain
\begin{align}
H^I\approx \frac{\Omega_1^d\eta_1}{2}\left(i\sigma_+\adag e^{i\delta_1 t}e^{-i\phi_1^d}+\textrm{H.c.}\right)+\frac{\Omega_2^d\eta_2}{2}\left(i\sigma_+a e^{i\delta_2 t}e^{-i\phi_2^d}+\textrm{H.c.}\right). 
\end{align}
We choose $\Omega_1^d=\Omega_2^d=\Omega^d$ and $\eta_1=\eta_2=\eta$, which leads to
\begin{align}
H^I&\approx \frac{\Omega^d\eta}{2}\left(\sigma_+\left[iae^{i\delta_2 t}e^{-i\phi_2^d}+i\adag e^{i\delta_1 t}e^{-i\phi_1^d}\right]+\textrm{H.c.}\right)\\
&\approx -\frac{\Omega^d\eta}{2}\left(\sigma_+e^{i\tilde{\Omega}t}+\sigma_-e^{-i\tilde{\Omega}t}\right)\left(ae^{-i\tilde{\omega}_0t}+\adag e^{i\tilde{\omega}_0t}\right)
\end{align}
where in the last step the phases $\phi_{1,2}^d=3\pi/2$ were introduced, and $2\tilde{\Omega}=\delta_1+\delta_2$ as well as $2\tilde{\omega}_0=\delta_1-\delta_2$. Note that the previous Hamiltonian adopts the form of a Rabi model in the rotating frame of $H=\tilde{\Omega}/2 \sigma_z+\tilde{\omega}_0\adaga$,
\begin{align}
e^{-i\left(\tilde{\Omega}/2 \sigma_z+\tilde{\omega}_0\adaga\right)t}H^Ie^{i\left(\tilde{\Omega}/2 \sigma_z+\tilde{\omega}_0\adaga\right)t}\approx \tilde{\omega}_0\adaga +\frac{\tilde{\Omega}}{2}\sigma_z-\frac{\eta\Omega^d}{2}\sigma_x\left(a+\adag\right).
\end{align}

\begin{figure}[t]
\centering
\includegraphics[width=0.7\linewidth,angle=0]{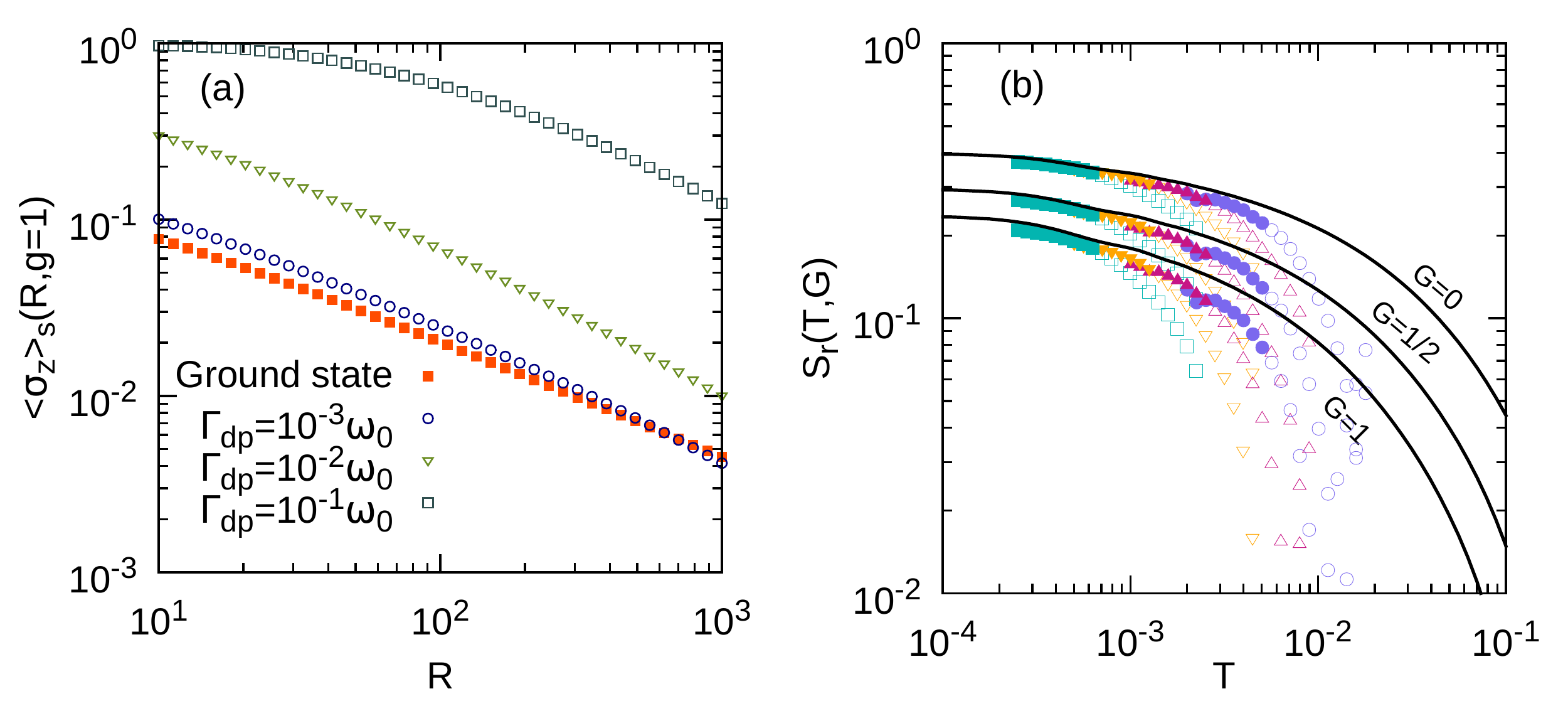}
\caption{In panel (a) the finite-frequency scaling of $\braket{\sigma_z}(R,g=1)$ is presented. The orange squares correspond to the ground state value, as in Fig.~\ref{fig:szScaling}, which follows the scaling $\braket{\sigma_z}(R,g=1)\propto R^{-\mu}$. The open points correspond to an adiabatic preparation of the ground state for $\tau_q=50/\omega_0$ with noises, $\Gamma_{dp}=10^{-1}\omega_0$, $10^{-2}\omega_0$ and $10^{-3}\omega_0$, being $\Gamma_{c,a,h}=\Gamma_{dp}/2$. The smaller the noise rates, the closer to the ground-state $\braket{\sigma_z}(R,g=1)$. Note, however, that the results do not follow the scaling $R^{-\mu}$, while making use of the finite-frequency scaling function $S_s(G)$ one can indeed witness the scaling exponent $\mu$ (see Fig.~3 (a) in the main text). In panel (b) we show the dynamical scaling function $S_r(T,G)$ for $\Gamma_{dp}=0.1\omega_0$ and $\Gamma_{c,a,h}=0.05\omega_0$, as in Fig.~3 (a), but including longer quench times. For $0.1\leq \omega_0\tau_q\leq 0.275$ (full points) the data follows reasonable well the predicted scaling function, while for  $0.275\leq \omega_0\tau_q\leq 1.0$ (open points) stronger deviations appear as a consequence of the noises, which deteriorate the universal curves.}
\label{fig:szDeph}
\end{figure}

In the main text, we have shown that the realization of the Rabi model with the traveling wave configuration cannot be used to observe the predicted universal properties of the Rabi model. This is mainly because the carrier interaction term, which has been neglected previously by vibrational RWA, becomes more relevant as the detunings $\delta_{1,2}$ increase. To resolve this issue, we consider a standing wave configuration for each sideband transition. More specifically the blue sideband is now driven by two traveling lasers, labeled with $1$ and $3$. The resulting Hamiltonian is
\begin{align}
H^I&\approx \frac{\Omega_1^d}{2}\left(\sigma_+e^{i((-\nu+\delta_1)t-\phi_1^d)}+\textrm{H.c}\right)+\frac{\Omega_3^d}{2}\left(\sigma_+e^{i((-\nu+\delta_3)t-\phi_3^d)}+\textrm{H.c}\right)+\nonumber \\
&+\frac{\Omega_1^d}{2}\left(i\eta_1\sigma_+\left(ae^{-i\nu t}+\adag e^{i\nu t}\right) e^{i((-\nu+\delta_1)t-\phi_1^d)}+\textrm{H.c}\right)+\frac{\Omega_3^d}{2}\left(i\eta_3\sigma_+\left(ae^{-i\nu t}+\adag e^{i\nu t}\right)e^{i((-\nu+\delta_3)t-\phi_3^d)}+\textrm{H.c}\right).
\end{align}
Therefore, by choosing $\Omega_1^d=\Omega_3^d$, $\delta_1=\delta_3$, $e^{-i\phi_1^d}=-e^{-i\phi_3^d}$, and $\eta_1=-\eta_3$, the carrier interaction term can be suppressed while the blue sideband interaction term remains. This configuration means that the two lasers are counter-propagating, realizing a standing wave, and the ion is placed in the node of the standing wave. The procedure to obtain the red detuned standing wave is identical. The parameters are then $\Omega_{1,2,3,4}^d=\Omega^d$, $\phi_{1,2}=3\pi/2$, $\phi_{3,4}=\pi/2$, $\omega_{1,3}^d=\omega_I+\nu-\delta_1$, $\omega_{2,4}^d=\omega_I-\nu-\delta_2$, $\eta_1=-\eta_3=\eta$ and $\eta_2=-\eta_4=\eta$, and the resulting Hamiltonian reads
\begin{align}
H^I\approx-\Omega^d\eta\left(\sigma_+e^{i\tilde{\Omega}t}+\sigma_-e^{-i\tilde{\Omega}t}\right)\left(ae^{-i\tilde{\omega}_0t}+\adag e^{i\tilde{\omega}_0t}\right),
\end{align}
with $\tilde{\Omega}$ and $\tilde{\omega}_0$ as defined previously. Note that the coupling strength is now the double as compared with the traveling wave setting.

The standing wave configuration may have an experimental limitation, namely, the intensities of the two counter-propagating lasers may be different as discussed in a recent experiment \cite{deLaubenfels:2015fbs}. Therefore, we analyze the impact of such experimental imperfection, $\Omega_1^d\neq\Omega_3^d$ and  $\Omega_2^d\neq\Omega_4^d$. For simplicity, we consider $\Omega_{3,4}^d=\Omega_f^d(1+\epsilon/2)$ and $\Omega_{3,4}^d=\Omega_f^d(1-\epsilon/2)$ where the $\Omega_f^d=(\Omega_{1,2}^d+\Omega_{3,4}^d)/2$ is the desired Rabi frequency and the error is set by $\epsilon$. Note that in Ref.~\cite{deLaubenfels:2015fbs}, $8\%$ ($\epsilon=0.08$) of error is present in the counter-propagating laser intensities. The results with $\epsilon=0.08$ are plotted in Fig.~2 (c) and (d) of the main text.

\section{Section E: Effect of noises}
\label{sec:E}
%tauq>0.275/w0 in Fig(b)
%Finite-size scaling at g_c for master equation: why the other approach is better.
In this section, we show the effect of noises for longer quench times $\tau_q$ which might deteriorate the universal scaling functions. The dynamics of the system is modeled by the following master equation
\begin{align}
\label{eq:me}
\dot{\rho}=-i\left[H_{\textrm{Rabi}}(t),\rho \right]+\Gamma_{dp}\mathcal{L}[\sigma_z]+\Gamma_{c}\mathcal{L}[\sigma_-]+\Gamma_{a}\mathcal{L}[a]+\Gamma_{h}\mathcal{L}[\adag],
\end{align}
where $\mathcal{L}[x]=x\rho x^{\dagger}-x^{\dagger}x\rho/2-\rho x^{\dagger}x/2$ is the Lindbladian superoperator, and $\Gamma$ is the corresponding rate. We consider noise rates for the trapped-ion setting as $\Gamma_{dp}=2\pi\times 20\Hz$, and $\Gamma_{c,a,h}=2\pi\times 10\Hz$, thus, $\Gamma_{dp}/\omega_0=0.1$ and $\Gamma_{c,a,h}/\omega_0=0.05$. However, one can study the impact of noise rates in the dynamics. For that, we choose $\Gamma_{dp}=10^{-1}\omega_0$, $10^{-2}\omega_0$ and $10^{-3}\omega_0$, being $\Gamma_{c,a,h}=\Gamma_{dp}/2$. Then, we can solve the dynamics governed by Eq.~(\ref{eq:me}) for different $R$ and quench times $\tau_q$. In the panel (a) of Fig.~\ref{fig:szDeph}, we plot the result of the adiabatic preparation of the ground state at the critical point for a fixed quench time $\omega_0\tau_q=50$ and different noise rates to witness the scaling relation $\braket{\sigma_z}_s(R,g=1)\propto R^{-\mu}$. The effect of the noises obscures the power-law scaling, and the scaling exponent $\mu$ is difficult to be measured this way. In the Fig.~3 (a) of the main text, however, we have shown that the asymptotic behavior of the scaling function is more robust to the effect of noises and that it allows one to measure the scaling exponent $\mu$ even in the presence of noises. In the panel (b) of Fig.~\ref{fig:szDeph}, we show the data for $S_r(T,G)$ with noise rates $\Gamma_{dp}=0.1\omega_0$ and $\Gamma_{c,a,h}=0.05\omega_0$ for quench times $0.1\leq \omega_0\tau_q\leq 0.275$ (full points) as in Fig.~3 (b), and  $0.275\leq \omega_0\tau_q\leq 1.0$ (open points). The dynamical scaling function $S_r(T,G)$ is deteriorated for longer quench times as the noises introduce excitations in the system which do not obey the universality of the QPT. Therefore, for these given noise rates, $S_r(T,G)$ remains intact when the quench time is in the range of $0.1\leq \omega_0\tau_q\leq 0.275$. Smaller noise rates would lead to a broader range of quench times $\tau_q$ for which  $S_r(T,G)$ can be observed.


\begin{thebibliography}{40}%
\makeatletter
\providecommand \@ifxundefined [1]{%
 \@ifx{#1\undefined}
}%
\providecommand \@ifnum [1]{%
 \ifnum #1\expandafter \@firstoftwo
 \else \expandafter \@secondoftwo
 \fi
}%
\providecommand \@ifx [1]{%
 \ifx #1\expandafter \@firstoftwo
 \else \expandafter \@secondoftwo
 \fi
}%
\providecommand \natexlab [1]{#1}%
\providecommand \enquote  [1]{``#1''}%
\providecommand \bibnamefont  [1]{#1}%
\providecommand \bibfnamefont [1]{#1}%
\providecommand \citenamefont [1]{#1}%
\providecommand \href@noop [0]{\@secondoftwo}%
\providecommand \href [0]{\begingroup \@sanitize@url \@href}%
\providecommand \@href[1]{\@@startlink{#1}\@@href}%
\providecommand \@@href[1]{\endgroup#1\@@endlink}%
\providecommand \@sanitize@url [0]{\catcode `\\12\catcode `\$12\catcode
  `\&12\catcode `\#12\catcode `\^12\catcode `\_12\catcode `\%12\relax}%
\providecommand \@@startlink[1]{}%
\providecommand \@@endlink[0]{}%
\providecommand \url  [0]{\begingroup\@sanitize@url \@url }%
\providecommand \@url [1]{\endgroup\@href {#1}{\urlprefix }}%
\providecommand \urlprefix  [0]{URL }%
\providecommand \Eprint [0]{\href }%
\providecommand \doibase [0]{http://dx.doi.org/}%
\providecommand \selectlanguage [0]{\@gobble}%
\providecommand \bibinfo  [0]{\@secondoftwo}%
\providecommand \bibfield  [0]{\@secondoftwo}%
\providecommand \translation [1]{[#1]}%
\providecommand \BibitemOpen [0]{}%
\providecommand \bibitemStop [0]{}%
\providecommand \bibitemNoStop [0]{.\EOS\space}%
\providecommand \EOS [0]{\spacefactor3000\relax}%
\providecommand \BibitemShut  [1]{\csname bibitem#1\endcsname}%
\let\auto@bib@innerbib\@empty
%</preamble>
\bibitem [{\citenamefont {Greiner}\ \emph {et~al.}(2002)\citenamefont
  {Greiner}, \citenamefont {Mandel}, \citenamefont {Esslinger}, \citenamefont
  {H{\"a}nsch},\ and\ \citenamefont {Bloch}}]{Greiner:2002es}%
  \BibitemOpen
  \bibfield  {author} {\bibinfo {author} {\bibfnamefont {M.}~\bibnamefont
  {Greiner}}, \bibinfo {author} {\bibfnamefont {O.}~\bibnamefont {Mandel}},
  \bibinfo {author} {\bibfnamefont {T.}~\bibnamefont {Esslinger}}, \bibinfo
  {author} {\bibfnamefont {T.~W.}\ \bibnamefont {H{\"a}nsch}}, \ and\ \bibinfo
  {author} {\bibfnamefont {I.}~\bibnamefont {Bloch}},\ }\href@noop {}
  {\bibfield  {journal} {\bibinfo  {journal} {Nature}\ }\textbf {\bibinfo
  {volume} {415}},\ \bibinfo {pages} {39} (\bibinfo {year} {2002})}\BibitemShut
  {NoStop}%
\bibitem [{\citenamefont {Bloch}\ \emph {et~al.}(2012)\citenamefont {Bloch},
  \citenamefont {Dalibard},\ and\ \citenamefont
  {Nascimb{\`e}ne}}]{Bloch:2012jy}%
  \BibitemOpen
  \bibfield  {author} {\bibinfo {author} {\bibfnamefont {I.}~\bibnamefont
  {Bloch}}, \bibinfo {author} {\bibfnamefont {J.}~\bibnamefont {Dalibard}}, \
  and\ \bibinfo {author} {\bibfnamefont {S.}~\bibnamefont {Nascimb{\`e}ne}},\
  }\href@noop {} {\bibfield  {journal} {\bibinfo  {journal} {Nature Physics}\
  }\textbf {\bibinfo {volume} {8}},\ \bibinfo {pages} {267} (\bibinfo {year}
  {2012})}\BibitemShut {NoStop}%
\bibitem [{\citenamefont {Porras}\ and\ \citenamefont
  {Cirac}(2004)}]{Porras:2004di}%
  \BibitemOpen
  \bibfield  {author} {\bibinfo {author} {\bibfnamefont {D.}~\bibnamefont
  {Porras}}\ and\ \bibinfo {author} {\bibfnamefont {J.~I.}\ \bibnamefont
  {Cirac}},\ }\href@noop {} {\bibfield  {journal} {\bibinfo  {journal}
  {Physical Review Letters}\ }\textbf {\bibinfo {volume} {92}},\ \bibinfo
  {pages} {207901} (\bibinfo {year} {2004})}\BibitemShut {NoStop}%
\bibitem [{\citenamefont {Kim}\ \emph {et~al.}(2010)\citenamefont {Kim},
  \citenamefont {Chang}, \citenamefont {Korenblit}, \citenamefont {Islam},
  \citenamefont {Edwards}, \citenamefont {Freericks}, \citenamefont {Lin},
  \citenamefont {Duan},\ and\ \citenamefont {Monroe}}]{Kim:2010ib}%
  \BibitemOpen
  \bibfield  {author} {\bibinfo {author} {\bibfnamefont {K.}~\bibnamefont
  {Kim}}, \bibinfo {author} {\bibfnamefont {M.~S.}\ \bibnamefont {Chang}},
  \bibinfo {author} {\bibfnamefont {S.}~\bibnamefont {Korenblit}}, \bibinfo
  {author} {\bibfnamefont {R.}~\bibnamefont {Islam}}, \bibinfo {author}
  {\bibfnamefont {E.~E.}\ \bibnamefont {Edwards}}, \bibinfo {author}
  {\bibfnamefont {J.~K.}\ \bibnamefont {Freericks}}, \bibinfo {author}
  {\bibfnamefont {G.~D.}\ \bibnamefont {Lin}}, \bibinfo {author} {\bibfnamefont
  {L.~M.}\ \bibnamefont {Duan}}, \ and\ \bibinfo {author} {\bibfnamefont
  {C.}~\bibnamefont {Monroe}},\ }\href@noop {} {\bibfield  {journal} {\bibinfo
  {journal} {Nature}\ }\textbf {\bibinfo {volume} {465}},\ \bibinfo {pages}
  {590} (\bibinfo {year} {2010})}\BibitemShut {NoStop}%
\bibitem [{\citenamefont {Islam}\ \emph {et~al.}(2011)\citenamefont {Islam},
  \citenamefont {Edwards}, \citenamefont {Kim}, \citenamefont {Korenblit},
  \citenamefont {Noh}, \citenamefont {Carmichael}, \citenamefont {Lin},
  \citenamefont {Duan}, \citenamefont {Wang}, \citenamefont {Freericks},\ and\
  \citenamefont {Monroe}}]{Islam:2011ct}%
  \BibitemOpen
  \bibfield  {author} {\bibinfo {author} {\bibfnamefont {R.}~\bibnamefont
  {Islam}}, \bibinfo {author} {\bibfnamefont {E.~E.}\ \bibnamefont {Edwards}},
  \bibinfo {author} {\bibfnamefont {K.}~\bibnamefont {Kim}}, \bibinfo {author}
  {\bibfnamefont {S.}~\bibnamefont {Korenblit}}, \bibinfo {author}
  {\bibfnamefont {C.}~\bibnamefont {Noh}}, \bibinfo {author} {\bibfnamefont
  {H.}~\bibnamefont {Carmichael}}, \bibinfo {author} {\bibfnamefont {G.~D.}\
  \bibnamefont {Lin}}, \bibinfo {author} {\bibfnamefont {L.~M.}\ \bibnamefont
  {Duan}}, \bibinfo {author} {\bibfnamefont {C.~C.~J.}\ \bibnamefont {Wang}},
  \bibinfo {author} {\bibfnamefont {J.~K.}\ \bibnamefont {Freericks}}, \ and\
  \bibinfo {author} {\bibfnamefont {C.}~\bibnamefont {Monroe}},\ }\href@noop {}
  {\bibfield  {journal} {\bibinfo  {journal} {Nature Communications}\ }\textbf
  {\bibinfo {volume} {2}},\ \bibinfo {pages} {377} (\bibinfo {year}
  {2011})}\BibitemShut {NoStop}%
\bibitem [{\citenamefont {Bermudez}\ and\ \citenamefont
  {Plenio}(2012)}]{Bermudez:2012hs}%
  \BibitemOpen
  \bibfield  {author} {\bibinfo {author} {\bibfnamefont {A.}~\bibnamefont
  {Bermudez}}\ and\ \bibinfo {author} {\bibfnamefont {M.~B.}\ \bibnamefont
  {Plenio}},\ }\href@noop {} {\bibfield  {journal} {\bibinfo  {journal}
  {Physical Review Letters}\ }\textbf {\bibinfo {volume} {109}},\ \bibinfo
  {pages} {010501} (\bibinfo {year} {2012})}\BibitemShut {NoStop}%
\bibitem [{\citenamefont {Baumann}\ \emph {et~al.}(2010)\citenamefont
  {Baumann}, \citenamefont {Guerlin}, \citenamefont {Brennecke},\ and\
  \citenamefont {Esslinger}}]{Baumann:2010js}%
  \BibitemOpen
  \bibfield  {author} {\bibinfo {author} {\bibfnamefont {K.}~\bibnamefont
  {Baumann}}, \bibinfo {author} {\bibfnamefont {C.}~\bibnamefont {Guerlin}},
  \bibinfo {author} {\bibfnamefont {F.}~\bibnamefont {Brennecke}}, \ and\
  \bibinfo {author} {\bibfnamefont {T.}~\bibnamefont {Esslinger}},\ }\href@noop
  {} {\bibfield  {journal} {\bibinfo  {journal} {Nature}\ }\textbf {\bibinfo
  {volume} {464}},\ \bibinfo {pages} {1301} (\bibinfo {year}
  {2010})}\BibitemShut {NoStop}%
\bibitem [{\citenamefont {Dimer}\ \emph {et~al.}(2007)\citenamefont {Dimer},
  \citenamefont {Estienne}, \citenamefont {Parkins},\ and\ \citenamefont
  {Carmichael}}]{Dimer:2007da}%
  \BibitemOpen
  \bibfield  {author} {\bibinfo {author} {\bibfnamefont {F.}~\bibnamefont
  {Dimer}}, \bibinfo {author} {\bibfnamefont {B.}~\bibnamefont {Estienne}},
  \bibinfo {author} {\bibfnamefont {A.}~\bibnamefont {Parkins}}, \ and\
  \bibinfo {author} {\bibfnamefont {H.}~\bibnamefont {Carmichael}},\
  }\href@noop {} {\bibfield  {journal} {\bibinfo  {journal} {Physical Review
  A}\ }\textbf {\bibinfo {volume} {75}},\ \bibinfo {pages} {013804} (\bibinfo
  {year} {2007})}\BibitemShut {NoStop}%
\bibitem [{\citenamefont {Nagy}\ \emph {et~al.}(2010)\citenamefont {Nagy},
  \citenamefont {K{\'o}nya}, \citenamefont {Szirmai},\ and\ \citenamefont
  {Domokos}}]{Nagy:2010dr}%
  \BibitemOpen
  \bibfield  {author} {\bibinfo {author} {\bibfnamefont {D.}~\bibnamefont
  {Nagy}}, \bibinfo {author} {\bibfnamefont {G.}~\bibnamefont {K{\'o}nya}},
  \bibinfo {author} {\bibfnamefont {G.}~\bibnamefont {Szirmai}}, \ and\
  \bibinfo {author} {\bibfnamefont {P.}~\bibnamefont {Domokos}},\ }\href@noop
  {} {\bibfield  {journal} {\bibinfo  {journal} {Physical Review Letters}\
  }\textbf {\bibinfo {volume} {104}},\ \bibinfo {pages} {130401} (\bibinfo
  {year} {2010})}\BibitemShut {NoStop}%
\bibitem [{\citenamefont {Damski}(2005)}]{Damski:2005cr}%
  \BibitemOpen
  \bibfield  {author} {\bibinfo {author} {\bibfnamefont {B.}~\bibnamefont
  {Damski}},\ }\href@noop {} {\bibfield  {journal} {\bibinfo  {journal}
  {Physical Review Letters}\ }\textbf {\bibinfo {volume} {95}},\ \bibinfo
  {pages} {035701} (\bibinfo {year} {2005})}\BibitemShut {NoStop}%
\bibitem [{\citenamefont {Zurek}\ \emph {et~al.}(2005)\citenamefont {Zurek},
  \citenamefont {Dorner},\ and\ \citenamefont {Zoller}}]{Zurek:2005cu}%
  \BibitemOpen
  \bibfield  {author} {\bibinfo {author} {\bibfnamefont {W.}~\bibnamefont
  {Zurek}}, \bibinfo {author} {\bibfnamefont {U.}~\bibnamefont {Dorner}}, \
  and\ \bibinfo {author} {\bibfnamefont {P.}~\bibnamefont {Zoller}},\
  }\href@noop {} {\bibfield  {journal} {\bibinfo  {journal} {Physical Review
  Letters}\ }\textbf {\bibinfo {volume} {95}},\ \bibinfo {pages} {105701}
  (\bibinfo {year} {2005})}\BibitemShut {NoStop}%
\bibitem [{\citenamefont {Polkovnikov}(2005)}]{Polkovnikov:2005gr}%
  \BibitemOpen
  \bibfield  {author} {\bibinfo {author} {\bibfnamefont {A.}~\bibnamefont
  {Polkovnikov}},\ }\href@noop {} {\bibfield  {journal} {\bibinfo  {journal}
  {Physical Review B}\ }\textbf {\bibinfo {volume} {72}},\ \bibinfo {pages}
  {161201} (\bibinfo {year} {2005})}\BibitemShut {NoStop}%
\bibitem [{\citenamefont {Braun}\ \emph {et~al.}(2015)\citenamefont {Braun},
  \citenamefont {Friesdorf}, \citenamefont {Hodgman}, \citenamefont
  {Schreiber}, \citenamefont {Ronzheimer}, \citenamefont {Riera}, \citenamefont
  {del Rey}, \citenamefont {Bloch}, \citenamefont {Eisert},\ and\ \citenamefont
  {Schneider}}]{Braun:2015iw}%
  \BibitemOpen
  \bibfield  {author} {\bibinfo {author} {\bibfnamefont {S.}~\bibnamefont
  {Braun}}, \bibinfo {author} {\bibfnamefont {M.}~\bibnamefont {Friesdorf}},
  \bibinfo {author} {\bibfnamefont {S.~S.}\ \bibnamefont {Hodgman}}, \bibinfo
  {author} {\bibfnamefont {M.}~\bibnamefont {Schreiber}}, \bibinfo {author}
  {\bibfnamefont {J.~P.}\ \bibnamefont {Ronzheimer}}, \bibinfo {author}
  {\bibfnamefont {A.}~\bibnamefont {Riera}}, \bibinfo {author} {\bibfnamefont
  {M.}~\bibnamefont {del Rey}}, \bibinfo {author} {\bibfnamefont
  {I.}~\bibnamefont {Bloch}}, \bibinfo {author} {\bibfnamefont
  {J.}~\bibnamefont {Eisert}}, \ and\ \bibinfo {author} {\bibfnamefont
  {U.}~\bibnamefont {Schneider}},\ }\href@noop {} {\bibfield  {journal}
  {\bibinfo  {journal} {Proceedings of the National Academy of Sciences of the
  United States of America}\ }\textbf {\bibinfo {volume} {112}},\ \bibinfo
  {pages} {3641} (\bibinfo {year} {2015})}\BibitemShut {NoStop}%
\bibitem [{\citenamefont {Klinder}\ \emph {et~al.}(2015)\citenamefont
  {Klinder}, \citenamefont {Ke{\ss}ler}, \citenamefont {Wolke}, \citenamefont
  {Mathey},\ and\ \citenamefont {Hemmerich}}]{Klinder:2015df}%
  \BibitemOpen
  \bibfield  {author} {\bibinfo {author} {\bibfnamefont {J.}~\bibnamefont
  {Klinder}}, \bibinfo {author} {\bibfnamefont {H.}~\bibnamefont {Ke{\ss}ler}},
  \bibinfo {author} {\bibfnamefont {M.}~\bibnamefont {Wolke}}, \bibinfo
  {author} {\bibfnamefont {L.}~\bibnamefont {Mathey}}, \ and\ \bibinfo {author}
  {\bibfnamefont {A.}~\bibnamefont {Hemmerich}},\ }\href@noop {} {\bibfield
  {journal} {\bibinfo  {journal} {Proceedings of the National Academy of
  Sciences of the United States of America}\ }\textbf {\bibinfo {volume}
  {112}},\ \bibinfo {pages} {3290} (\bibinfo {year} {2015})}\BibitemShut
  {NoStop}%
\bibitem [{\citenamefont {Polkovnikov}\ \emph {et~al.}(2011)\citenamefont
  {Polkovnikov}, \citenamefont {Sengupta}, \citenamefont {Silva},\ and\
  \citenamefont {Vengalattore}}]{Polkovnikov:2011iu}%
  \BibitemOpen
  \bibfield  {author} {\bibinfo {author} {\bibfnamefont {A.}~\bibnamefont
  {Polkovnikov}}, \bibinfo {author} {\bibfnamefont {K.}~\bibnamefont
  {Sengupta}}, \bibinfo {author} {\bibfnamefont {A.}~\bibnamefont {Silva}}, \
  and\ \bibinfo {author} {\bibfnamefont {M.}~\bibnamefont {Vengalattore}},\
  }\href@noop {} {\bibfield  {journal} {\bibinfo  {journal} {Reviews of Modern
  Physics}\ }\textbf {\bibinfo {volume} {83}},\ \bibinfo {pages} {863}
  (\bibinfo {year} {2011})}\BibitemShut {NoStop}%
\bibitem [{\citenamefont {Eisert}\ \emph {et~al.}(2015)\citenamefont {Eisert},
  \citenamefont {Friesdorf},\ and\ \citenamefont {Gogolin}}]{Eisert:2015ka}%
  \BibitemOpen
  \bibfield  {author} {\bibinfo {author} {\bibfnamefont {J.}~\bibnamefont
  {Eisert}}, \bibinfo {author} {\bibfnamefont {M.}~\bibnamefont {Friesdorf}}, \
  and\ \bibinfo {author} {\bibfnamefont {C.}~\bibnamefont {Gogolin}},\
  }\href@noop {} {\bibfield  {journal} {\bibinfo  {journal} {Nature Physics}\
  }\textbf {\bibinfo {volume} {11}},\ \bibinfo {pages} {124} (\bibinfo {year}
  {2015})}\BibitemShut {NoStop}%
\bibitem [{\citenamefont {Kolodrubetz}\ \emph {et~al.}(2012)\citenamefont
  {Kolodrubetz}, \citenamefont {Clark},\ and\ \citenamefont
  {Huse}}]{Kolodrubetz:2012jo}%
  \BibitemOpen
  \bibfield  {author} {\bibinfo {author} {\bibfnamefont {M.}~\bibnamefont
  {Kolodrubetz}}, \bibinfo {author} {\bibfnamefont {B.~K.}\ \bibnamefont
  {Clark}}, \ and\ \bibinfo {author} {\bibfnamefont {D.~A.}\ \bibnamefont
  {Huse}},\ }\href@noop {} {\bibfield  {journal} {\bibinfo  {journal} {Physical
  Review Letters}\ }\textbf {\bibinfo {volume} {109}},\ \bibinfo {pages}
  {015701} (\bibinfo {year} {2012})}\BibitemShut {NoStop}%
\bibitem [{\citenamefont {Nikoghosyan}\ \emph {et~al.}(2016)\citenamefont
  {Nikoghosyan}, \citenamefont {Nigmatullin},\ and\ \citenamefont
  {Plenio}}]{Nikoghosyan:2016ff}%
  \BibitemOpen
  \bibfield  {author} {\bibinfo {author} {\bibfnamefont {G.}~\bibnamefont
  {Nikoghosyan}}, \bibinfo {author} {\bibfnamefont {R.}~\bibnamefont
  {Nigmatullin}}, \ and\ \bibinfo {author} {\bibfnamefont {M.~B.}\ \bibnamefont
  {Plenio}},\ }\href@noop {} {\bibfield  {journal} {\bibinfo  {journal}
  {Physical Review Letters}\ }\textbf {\bibinfo {volume} {116}},\ \bibinfo
  {pages} {080601} (\bibinfo {year} {2016})}\BibitemShut {NoStop}%
\bibitem [{\citenamefont {Acevedo}\ \emph {et~al.}(2014)\citenamefont
  {Acevedo}, \citenamefont {Quiroga}, \citenamefont {Rodr{\'\i}guez},\ and\
  \citenamefont {Johnson}}]{Acevedo:2014eo}%
  \BibitemOpen
  \bibfield  {author} {\bibinfo {author} {\bibfnamefont {O.~L.}\ \bibnamefont
  {Acevedo}}, \bibinfo {author} {\bibfnamefont {L.}~\bibnamefont {Quiroga}},
  \bibinfo {author} {\bibfnamefont {F.~J.}\ \bibnamefont {Rodr{\'\i}guez}}, \
  and\ \bibinfo {author} {\bibfnamefont {N.~F.}\ \bibnamefont {Johnson}},\
  }\href@noop {} {\bibfield  {journal} {\bibinfo  {journal} {Physical Review
  Letters}\ }\textbf {\bibinfo {volume} {112}},\ \bibinfo {pages} {030403}
  (\bibinfo {year} {2014})}\BibitemShut {NoStop}%
\bibitem [{\citenamefont {Hwang}\ \emph {et~al.}(2015)\citenamefont {Hwang},
  \citenamefont {Puebla},\ and\ \citenamefont {Plenio}}]{Hwang:2015eq}%
  \BibitemOpen
  \bibfield  {author} {\bibinfo {author} {\bibfnamefont {M.-J.}\ \bibnamefont
  {Hwang}}, \bibinfo {author} {\bibfnamefont {R.}~\bibnamefont {Puebla}}, \
  and\ \bibinfo {author} {\bibfnamefont {M.~B.}\ \bibnamefont {Plenio}},\
  }\href@noop {} {\bibfield  {journal} {\bibinfo  {journal} {Physical Review
  Letters}\ }\textbf {\bibinfo {volume} {115}},\ \bibinfo {pages} {180404}
  (\bibinfo {year} {2015})}\BibitemShut {NoStop}%
\bibitem [{\citenamefont {Pyka}\ \emph {et~al.}(2013)\citenamefont {Pyka},
  \citenamefont {Keller}, \citenamefont {Partner}, \citenamefont {Nigmatullin},
  \citenamefont {Burgermeister}, \citenamefont {Meier}, \citenamefont
  {Kuhlmann}, \citenamefont {Retzker}, \citenamefont {Plenio}, \citenamefont
  {Zurek} \emph {et~al.}}]{pyka2013topological}%
  \BibitemOpen
  \bibfield  {author} {\bibinfo {author} {\bibfnamefont {K.}~\bibnamefont
  {Pyka}}, \bibinfo {author} {\bibfnamefont {J.}~\bibnamefont {Keller}},
  \bibinfo {author} {\bibfnamefont {H.}~\bibnamefont {Partner}}, \bibinfo
  {author} {\bibfnamefont {R.}~\bibnamefont {Nigmatullin}}, \bibinfo {author}
  {\bibfnamefont {T.}~\bibnamefont {Burgermeister}}, \bibinfo {author}
  {\bibfnamefont {D.}~\bibnamefont {Meier}}, \bibinfo {author} {\bibfnamefont
  {K.}~\bibnamefont {Kuhlmann}}, \bibinfo {author} {\bibfnamefont
  {A.}~\bibnamefont {Retzker}}, \bibinfo {author} {\bibfnamefont {M.~B.}\
  \bibnamefont {Plenio}}, \bibinfo {author} {\bibfnamefont {W.}~\bibnamefont
  {Zurek}},  \emph {et~al.},\ }\href@noop {} {\bibfield  {journal} {\bibinfo
  {journal} {Nature Communications}\ }\textbf {\bibinfo {volume} {4}},\
  \bibinfo {pages} {2291} (\bibinfo {year} {2013})}\BibitemShut {NoStop}%
\bibitem [{\citenamefont {Ulm}\ \emph {et~al.}(2013)\citenamefont {Ulm},
  \citenamefont {Ro{\ss}nagel}, \citenamefont {Jacob}, \citenamefont
  {Deg{\"u}nther}, \citenamefont {Dawkins}, \citenamefont {Poschinger},
  \citenamefont {Nigmatullin}, \citenamefont {Retzker}, \citenamefont {Plenio},
  \citenamefont {Schmidt-Kaler} \emph {et~al.}}]{ulm2013observation}%
  \BibitemOpen
  \bibfield  {author} {\bibinfo {author} {\bibfnamefont {S.}~\bibnamefont
  {Ulm}}, \bibinfo {author} {\bibfnamefont {J.}~\bibnamefont {Ro{\ss}nagel}},
  \bibinfo {author} {\bibfnamefont {G.}~\bibnamefont {Jacob}}, \bibinfo
  {author} {\bibfnamefont {C.}~\bibnamefont {Deg{\"u}nther}}, \bibinfo {author}
  {\bibfnamefont {S.}~\bibnamefont {Dawkins}}, \bibinfo {author} {\bibfnamefont
  {U.}~\bibnamefont {Poschinger}}, \bibinfo {author} {\bibfnamefont
  {R.}~\bibnamefont {Nigmatullin}}, \bibinfo {author} {\bibfnamefont
  {A.}~\bibnamefont {Retzker}}, \bibinfo {author} {\bibfnamefont {M.~B.}\
  \bibnamefont {Plenio}}, \bibinfo {author} {\bibfnamefont {F.}~\bibnamefont
  {Schmidt-Kaler}},  \emph {et~al.},\ }\href@noop {} {\bibfield  {journal}
  {\bibinfo  {journal} {Nature communications}\ }\textbf {\bibinfo {volume}
  {4}},\ \bibinfo {pages} {2290} (\bibinfo {year} {2013})}\BibitemShut
  {NoStop}%
\bibitem [{\citenamefont {Sachdev}(2011)}]{Sachdev:2011uj}%
  \BibitemOpen
  \bibfield  {author} {\bibinfo {author} {\bibfnamefont {S.}~\bibnamefont
  {Sachdev}},\ }\href@noop {} {\emph {\bibinfo {title} {{Quantum Phase
  Transitions}}}},\ \bibinfo {edition} {2nd}\ ed.\ (\bibinfo  {publisher}
  {Cambridge University Press},\ \bibinfo {year} {2011})\BibitemShut {NoStop}%
\bibitem [{\citenamefont {Fisher}\ and\ \citenamefont
  {Barber}(1972)}]{Fisher:1972tv}%
  \BibitemOpen
  \bibfield  {author} {\bibinfo {author} {\bibfnamefont {M.~E.}\ \bibnamefont
  {Fisher}}\ and\ \bibinfo {author} {\bibfnamefont {M.~N.}\ \bibnamefont
  {Barber}},\ }\href@noop {} {\bibfield  {journal} {\bibinfo  {journal}
  {Physical Review Letters}\ }\textbf {\bibinfo {volume} {28}},\ \bibinfo
  {pages} {1516} (\bibinfo {year} {1972})}\BibitemShut {NoStop}%
\bibitem [{\citenamefont {Botet}\ \emph {et~al.}(1982)\citenamefont {Botet},
  \citenamefont {Jullien},\ and\ \citenamefont {Pfeuty}}]{Botet:1982ju}%
  \BibitemOpen
  \bibfield  {author} {\bibinfo {author} {\bibfnamefont {R.}~\bibnamefont
  {Botet}}, \bibinfo {author} {\bibfnamefont {R.}~\bibnamefont {Jullien}}, \
  and\ \bibinfo {author} {\bibfnamefont {P.}~\bibnamefont {Pfeuty}},\
  }\href@noop {} {\bibfield  {journal} {\bibinfo  {journal} {Physical Review
  Letters}\ }\textbf {\bibinfo {volume} {49}},\ \bibinfo {pages} {478}
  (\bibinfo {year} {1982})}\BibitemShut {NoStop}%
\bibitem [{\citenamefont {Hwang}\ and\ \citenamefont
  {Plenio}(2016)}]{Hwang:2016vf}%
  \BibitemOpen
  \bibfield  {author} {\bibinfo {author} {\bibfnamefont {M.-J.}\ \bibnamefont
  {Hwang}}\ and\ \bibinfo {author} {\bibfnamefont {M.~B.}\ \bibnamefont
  {Plenio}},\ }\href@noop {} {\bibfield  {journal} {\bibinfo  {journal}
  {arXiv:1603.03943}\ } (\bibinfo {year} {2016})}\BibitemShut {NoStop}%
\bibitem [{\citenamefont {Ashhab}(2013)}]{Ashhab:2013ke}%
  \BibitemOpen
  \bibfield  {author} {\bibinfo {author} {\bibfnamefont {S.}~\bibnamefont
  {Ashhab}},\ }\href@noop {} {\bibfield  {journal} {\bibinfo  {journal}
  {Physical Review A}\ }\textbf {\bibinfo {volume} {87}},\ \bibinfo {pages}
  {013826} (\bibinfo {year} {2013})}\BibitemShut {NoStop}%
\bibitem [{\citenamefont {Bakemeier}\ \emph {et~al.}(2012)\citenamefont
  {Bakemeier}, \citenamefont {Alvermann},\ and\ \citenamefont
  {Fehske}}]{Bakemeier:2012ja}%
  \BibitemOpen
  \bibfield  {author} {\bibinfo {author} {\bibfnamefont {L.}~\bibnamefont
  {Bakemeier}}, \bibinfo {author} {\bibfnamefont {A.}~\bibnamefont
  {Alvermann}}, \ and\ \bibinfo {author} {\bibfnamefont {H.}~\bibnamefont
  {Fehske}},\ }\href@noop {} {\bibfield  {journal} {\bibinfo  {journal}
  {Physical Review A}\ }\textbf {\bibinfo {volume} {85}},\ \bibinfo {pages}
  {043821} (\bibinfo {year} {2012})}\BibitemShut {NoStop}%
\bibitem [{\citenamefont {Hwang}\ and\ \citenamefont
  {Choi}(2010)}]{Hwang:2010jn}%
  \BibitemOpen
  \bibfield  {author} {\bibinfo {author} {\bibfnamefont {M.-J.}\ \bibnamefont
  {Hwang}}\ and\ \bibinfo {author} {\bibfnamefont {M.-S.}\ \bibnamefont
  {Choi}},\ }\href@noop {} {\bibfield  {journal} {\bibinfo  {journal} {Physical
  Review A}\ }\textbf {\bibinfo {volume} {82}},\ \bibinfo {pages} {025802}
  (\bibinfo {year} {2010})}\BibitemShut {NoStop}%
\bibitem [{\citenamefont {Bourassa}\ \emph {et~al.}(2009)\citenamefont
  {Bourassa}, \citenamefont {Gambetta}, \citenamefont {Abdumalikov},
  \citenamefont {Astafiev}, \citenamefont {Nakamura},\ and\ \citenamefont
  {Blais}}]{Bourassa:2009gy}%
  \BibitemOpen
  \bibfield  {author} {\bibinfo {author} {\bibfnamefont {J.}~\bibnamefont
  {Bourassa}}, \bibinfo {author} {\bibfnamefont {J.}~\bibnamefont {Gambetta}},
  \bibinfo {author} {\bibfnamefont {A.}~\bibnamefont {Abdumalikov}}, \bibinfo
  {author} {\bibfnamefont {O.}~\bibnamefont {Astafiev}}, \bibinfo {author}
  {\bibfnamefont {Y.}~\bibnamefont {Nakamura}}, \ and\ \bibinfo {author}
  {\bibfnamefont {A.}~\bibnamefont {Blais}},\ }\href@noop {} {\bibfield
  {journal} {\bibinfo  {journal} {Physical Review A}\ }\textbf {\bibinfo
  {volume} {80}},\ \bibinfo {pages} {032109} (\bibinfo {year}
  {2009})}\BibitemShut {NoStop}%
\bibitem [{\citenamefont {Ashhab}\ and\ \citenamefont
  {Nori}(2010)}]{Ashhab:2010eh}%
  \BibitemOpen
  \bibfield  {author} {\bibinfo {author} {\bibfnamefont {S.}~\bibnamefont
  {Ashhab}}\ and\ \bibinfo {author} {\bibfnamefont {F.}~\bibnamefont {Nori}},\
  }\href@noop {} {\bibfield  {journal} {\bibinfo  {journal} {Physical Review
  A}\ }\textbf {\bibinfo {volume} {81}},\ \bibinfo {pages} {042311} (\bibinfo
  {year} {2010})}\BibitemShut {NoStop}%
\bibitem [{\citenamefont {Casanova}\ \emph {et~al.}(2010)\citenamefont
  {Casanova}, \citenamefont {Romero}, \citenamefont {Lizuain}, \citenamefont
  {Garc{\'\i}a-Ripoll},\ and\ \citenamefont {Solano}}]{Casanova:2010kd}%
  \BibitemOpen
  \bibfield  {author} {\bibinfo {author} {\bibfnamefont {J.}~\bibnamefont
  {Casanova}}, \bibinfo {author} {\bibfnamefont {G.}~\bibnamefont {Romero}},
  \bibinfo {author} {\bibfnamefont {I.}~\bibnamefont {Lizuain}}, \bibinfo
  {author} {\bibfnamefont {J.}~\bibnamefont {Garc{\'\i}a-Ripoll}}, \ and\
  \bibinfo {author} {\bibfnamefont {E.}~\bibnamefont {Solano}},\ }\href@noop {}
  {\bibfield  {journal} {\bibinfo  {journal} {Physical Review Letters}\
  }\textbf {\bibinfo {volume} {105}},\ \bibinfo {pages} {263603} (\bibinfo
  {year} {2010})}\BibitemShut {NoStop}%
\bibitem [{\citenamefont {Cirac}\ \emph {et~al.}(1993)\citenamefont {Cirac},
  \citenamefont {Parkins}, \citenamefont {Blatt},\ and\ \citenamefont
  {Zoller}}]{Cirac:1993db}%
  \BibitemOpen
  \bibfield  {author} {\bibinfo {author} {\bibfnamefont {J.~I.}\ \bibnamefont
  {Cirac}}, \bibinfo {author} {\bibfnamefont {A.~S.}\ \bibnamefont {Parkins}},
  \bibinfo {author} {\bibfnamefont {R.}~\bibnamefont {Blatt}}, \ and\ \bibinfo
  {author} {\bibfnamefont {P.}~\bibnamefont {Zoller}},\ }\href@noop {}
  {\bibfield  {journal} {\bibinfo  {journal} {Physical Review Letters}\
  }\textbf {\bibinfo {volume} {70}},\ \bibinfo {pages} {556} (\bibinfo {year}
  {1993})}\BibitemShut {NoStop}%
\bibitem [{\citenamefont {Pedernales}\ \emph {et~al.}(2015)\citenamefont
  {Pedernales}, \citenamefont {Lizuain}, \citenamefont {Felicetti},
  \citenamefont {Romero}, \citenamefont {Lamata},\ and\ \citenamefont
  {Solano}}]{Pedernales:2015cj}%
  \BibitemOpen
  \bibfield  {author} {\bibinfo {author} {\bibfnamefont {J.~S.}\ \bibnamefont
  {Pedernales}}, \bibinfo {author} {\bibfnamefont {I.}~\bibnamefont {Lizuain}},
  \bibinfo {author} {\bibfnamefont {S.}~\bibnamefont {Felicetti}}, \bibinfo
  {author} {\bibfnamefont {G.}~\bibnamefont {Romero}}, \bibinfo {author}
  {\bibfnamefont {L.}~\bibnamefont {Lamata}}, \ and\ \bibinfo {author}
  {\bibfnamefont {E.}~\bibnamefont {Solano}},\ }\href@noop {} {\bibfield
  {journal} {\bibinfo  {journal} {Scientific Reports}\ }\textbf {\bibinfo
  {volume} {5}},\ \bibinfo {pages} {15472} (\bibinfo {year}
  {2015})}\BibitemShut {NoStop}%
\bibitem [{\citenamefont {Cirac}\ \emph {et~al.}(1992)\citenamefont {Cirac},
  \citenamefont {Blatt}, \citenamefont {Zoller},\ and\ \citenamefont
  {Phillips}}]{Cirac:1992ck}%
  \BibitemOpen
  \bibfield  {author} {\bibinfo {author} {\bibfnamefont {J.~I.}\ \bibnamefont
  {Cirac}}, \bibinfo {author} {\bibfnamefont {R.}~\bibnamefont {Blatt}},
  \bibinfo {author} {\bibfnamefont {P.}~\bibnamefont {Zoller}}, \ and\ \bibinfo
  {author} {\bibfnamefont {W.~D.}\ \bibnamefont {Phillips}},\ }\href@noop {}
  {\bibfield  {journal} {\bibinfo  {journal} {Physical Review A}\ }\textbf
  {\bibinfo {volume} {46}},\ \bibinfo {pages} {2668} (\bibinfo {year}
  {1992})}\BibitemShut {NoStop}%
\bibitem [{\citenamefont {deLaubenfels}\ \emph {et~al.}(2015)\citenamefont
  {deLaubenfels}, \citenamefont {Burkhardt}, \citenamefont {Vittorini},
  \citenamefont {Merrill}, \citenamefont {Brown},\ and\ \citenamefont
  {Amini}}]{deLaubenfels:2015fb}%
  \BibitemOpen
  \bibfield  {author} {\bibinfo {author} {\bibfnamefont {T.~E.}\ \bibnamefont
  {deLaubenfels}}, \bibinfo {author} {\bibfnamefont {K.~A.}\ \bibnamefont
  {Burkhardt}}, \bibinfo {author} {\bibfnamefont {G.}~\bibnamefont
  {Vittorini}}, \bibinfo {author} {\bibfnamefont {J.~T.}\ \bibnamefont
  {Merrill}}, \bibinfo {author} {\bibfnamefont {K.~R.}\ \bibnamefont {Brown}},
  \ and\ \bibinfo {author} {\bibfnamefont {J.~M.}\ \bibnamefont {Amini}},\
  }\href@noop {} {\bibfield  {journal} {\bibinfo  {journal} {Physical Review
  A}\ }\textbf {\bibinfo {volume} {92}},\ \bibinfo {pages} {061402} (\bibinfo
  {year} {2015})}\BibitemShut {NoStop}%
\bibitem [{\citenamefont {Myerson}\ \emph {et~al.}(2008)\citenamefont
  {Myerson}, \citenamefont {Szwer}, \citenamefont {Webster}, \citenamefont
  {Allcock}, \citenamefont {Curtis}, \citenamefont {Imreh}, \citenamefont
  {Sherman}, \citenamefont {Stacey}, \citenamefont {Steane},\ and\
  \citenamefont {Lucas}}]{Myerson:2008bm}%
  \BibitemOpen
  \bibfield  {author} {\bibinfo {author} {\bibfnamefont {A.~H.}\ \bibnamefont
  {Myerson}}, \bibinfo {author} {\bibfnamefont {D.~J.}\ \bibnamefont {Szwer}},
  \bibinfo {author} {\bibfnamefont {S.~C.}\ \bibnamefont {Webster}}, \bibinfo
  {author} {\bibfnamefont {D.~T.~C.}\ \bibnamefont {Allcock}}, \bibinfo
  {author} {\bibfnamefont {M.~J.}\ \bibnamefont {Curtis}}, \bibinfo {author}
  {\bibfnamefont {G.}~\bibnamefont {Imreh}}, \bibinfo {author} {\bibfnamefont
  {J.~A.}\ \bibnamefont {Sherman}}, \bibinfo {author} {\bibfnamefont {D.~N.}\
  \bibnamefont {Stacey}}, \bibinfo {author} {\bibfnamefont {A.~M.}\
  \bibnamefont {Steane}}, \ and\ \bibinfo {author} {\bibfnamefont {D.~M.}\
  \bibnamefont {Lucas}},\ }\href@noop {} {\bibfield  {journal} {\bibinfo
  {journal} {Physical Review Letters}\ }\textbf {\bibinfo {volume} {100}},\
  \bibinfo {pages} {200502} (\bibinfo {year} {2008})}\BibitemShut {NoStop}%
\bibitem [{\citenamefont {Burrell}\ \emph {et~al.}(2010)\citenamefont
  {Burrell}, \citenamefont {Szwer}, \citenamefont {Webster},\ and\
  \citenamefont {Lucas}}]{Burrell:2010kh}%
  \BibitemOpen
  \bibfield  {author} {\bibinfo {author} {\bibfnamefont {A.~H.}\ \bibnamefont
  {Burrell}}, \bibinfo {author} {\bibfnamefont {D.~J.}\ \bibnamefont {Szwer}},
  \bibinfo {author} {\bibfnamefont {S.~C.}\ \bibnamefont {Webster}}, \ and\
  \bibinfo {author} {\bibfnamefont {D.~M.}\ \bibnamefont {Lucas}},\ }\href@noop
  {} {\bibfield  {journal} {\bibinfo  {journal} {Physical Review A}\ }\textbf
  {\bibinfo {volume} {81}},\ \bibinfo {pages} {040302} (\bibinfo {year}
  {2010})}\BibitemShut {NoStop}%
\bibitem [{sup()}]{sup}%
  \BibitemOpen
  \bibinfo {note} {See Supplemental Material at [URL will be inserted by
  publisher] for further explanation and details of the
  calculation}\BibitemShut {NoStop}%
\bibitem [{\citenamefont {Messiah}(1961)}]{Messiah:1961vw}%
  \BibitemOpen
  \bibfield  {author} {\bibinfo {author} {\bibfnamefont {A.}~\bibnamefont
  {Messiah}},\ }\href@noop {} {\emph {\bibinfo {title} {{Quantum Mechanics}}}}\
  (\bibinfo  {publisher} {Dover Publications},\ \bibinfo {year}
  {1961})\BibitemShut {NoStop}%
\end{thebibliography}

\begin{thebibliography}{7}%
\makeatletter
\providecommand \@ifxundefined [1]{%
 \@ifx{#1\undefined}
}%
\providecommand \@ifnum [1]{%
 \ifnum #1\expandafter \@firstoftwo
 \else \expandafter \@secondoftwo
 \fi
}%
\providecommand \@ifx [1]{%
 \ifx #1\expandafter \@firstoftwo
 \else \expandafter \@secondoftwo
 \fi
}%
\providecommand \natexlab [1]{#1}%
\providecommand \enquote  [1]{``#1''}%
\providecommand \bibnamefont  [1]{#1}%
\providecommand \bibfnamefont [1]{#1}%
\providecommand \citenamefont [1]{#1}%
\providecommand \href@noop [0]{\@secondoftwo}%
\providecommand \href [0]{\begingroup \@sanitize@url \@href}%
\providecommand \@href[1]{\@@startlink{#1}\@@href}%
\providecommand \@@href[1]{\endgroup#1\@@endlink}%
\providecommand \@sanitize@url [0]{\catcode `\\12\catcode `\$12\catcode
  `\&12\catcode `\#12\catcode `\^12\catcode `\_12\catcode `\%12\relax}%
\providecommand \@@startlink[1]{}%
\providecommand \@@endlink[0]{}%
\providecommand \url  [0]{\begingroup\@sanitize@url \@url }%
\providecommand \@url [1]{\endgroup\@href {#1}{\urlprefix }}%
\providecommand \urlprefix  [0]{URL }%
\providecommand \Eprint [0]{\href }%
\providecommand \doibase [0]{http://dx.doi.org/}%
\providecommand \selectlanguage [0]{\@gobble}%
\providecommand \bibinfo  [0]{\@secondoftwo}%
\providecommand \bibfield  [0]{\@secondoftwo}%
\providecommand \translation [1]{[#1]}%
\providecommand \BibitemOpen [0]{}%
\providecommand \bibitemStop [0]{}%
\providecommand \bibitemNoStop [0]{.\EOS\space}%
\providecommand \EOS [0]{\spacefactor3000\relax}%
\providecommand \BibitemShut  [1]{\csname bibitem#1\endcsname}%
\let\auto@bib@innerbib\@empty
%</preamble>
\bibitem [{\citenamefont {Hwang}\ \emph {et~al.}(2015)\citenamefont {Hwang},
  \citenamefont {Puebla},\ and\ \citenamefont {Plenio}}]{Hwang:2015eqs}%
  \BibitemOpen
  \bibfield  {author} {\bibinfo {author} {\bibfnamefont {M.-J.}\ \bibnamefont
  {Hwang}}, \bibinfo {author} {\bibfnamefont {R.}~\bibnamefont {Puebla}}, \
  and\ \bibinfo {author} {\bibfnamefont {M.~B.}\ \bibnamefont {Plenio}},\
  }\href@noop {} {\bibfield  {journal} {\bibinfo  {journal} {Physical Review
  Letters}\ }\textbf {\bibinfo {volume} {115}},\ \bibinfo {pages} {180404}
  (\bibinfo {year} {2015})}\BibitemShut {NoStop}%
\bibitem [{\citenamefont {Acevedo}\ \emph {et~al.}(2014)\citenamefont
  {Acevedo}, \citenamefont {Quiroga}, \citenamefont {Rodr{\'\i}guez},\ and\
  \citenamefont {Johnson}}]{Acevedo:2014eos}%
  \BibitemOpen
  \bibfield  {author} {\bibinfo {author} {\bibfnamefont {O.~L.}\ \bibnamefont
  {Acevedo}}, \bibinfo {author} {\bibfnamefont {L.}~\bibnamefont {Quiroga}},
  \bibinfo {author} {\bibfnamefont {F.~J.}\ \bibnamefont {Rodr{\'\i}guez}}, \
  and\ \bibinfo {author} {\bibfnamefont {N.~F.}\ \bibnamefont {Johnson}},\
  }\href@noop {} {\bibfield  {journal} {\bibinfo  {journal} {Physical Review
  Letters}\ }\textbf {\bibinfo {volume} {112}},\ \bibinfo {pages} {030403}
  (\bibinfo {year} {2014})}\BibitemShut {NoStop}%
\bibitem [{\citenamefont {Sachdev}(2011)}]{Sachdev:2011ujs}%
  \BibitemOpen
  \bibfield  {author} {\bibinfo {author} {\bibfnamefont {S.}~\bibnamefont
  {Sachdev}},\ }\href@noop {} {\emph {\bibinfo {title} {{Quantum Phase
  Transitions}}}},\ \bibinfo {edition} {2nd}\ ed.\ (\bibinfo  {publisher}
  {Cambridge University Press},\ \bibinfo {year} {2011})\BibitemShut {NoStop}%
\bibitem [{\citenamefont {Cirac}\ \emph {et~al.}(1993)\citenamefont {Cirac},
  \citenamefont {Parkins}, \citenamefont {Blatt},\ and\ \citenamefont
  {Zoller}}]{Cirac:1993dbs}%
  \BibitemOpen
  \bibfield  {author} {\bibinfo {author} {\bibfnamefont {J.~I.}\ \bibnamefont
  {Cirac}}, \bibinfo {author} {\bibfnamefont {A.~S.}\ \bibnamefont {Parkins}},
  \bibinfo {author} {\bibfnamefont {R.}~\bibnamefont {Blatt}}, \ and\ \bibinfo
  {author} {\bibfnamefont {P.}~\bibnamefont {Zoller}},\ }\href@noop {}
  {\bibfield  {journal} {\bibinfo  {journal} {Physical Review Letters}\
  }\textbf {\bibinfo {volume} {70}},\ \bibinfo {pages} {556} (\bibinfo {year}
  {1993})}\BibitemShut {NoStop}%
\bibitem [{\citenamefont {Pedernales}\ \emph {et~al.}(2015)\citenamefont
  {Pedernales}, \citenamefont {Lizuain}, \citenamefont {Felicetti},
  \citenamefont {Romero}, \citenamefont {Lamata},\ and\ \citenamefont
  {Solano}}]{Pedernales:2015cjs}%
  \BibitemOpen
  \bibfield  {author} {\bibinfo {author} {\bibfnamefont {J.~S.}\ \bibnamefont
  {Pedernales}}, \bibinfo {author} {\bibfnamefont {I.}~\bibnamefont {Lizuain}},
  \bibinfo {author} {\bibfnamefont {S.}~\bibnamefont {Felicetti}}, \bibinfo
  {author} {\bibfnamefont {G.}~\bibnamefont {Romero}}, \bibinfo {author}
  {\bibfnamefont {L.}~\bibnamefont {Lamata}}, \ and\ \bibinfo {author}
  {\bibfnamefont {E.}~\bibnamefont {Solano}},\ }\href@noop {} {\bibfield
  {journal} {\bibinfo  {journal} {Scientific Reports}\ }\textbf {\bibinfo
  {volume} {5}},\ \bibinfo {pages} {15472} (\bibinfo {year}
  {2015})}\BibitemShut {NoStop}%
\bibitem [{\citenamefont {Cirac}\ \emph {et~al.}(1992)\citenamefont {Cirac},
  \citenamefont {Blatt}, \citenamefont {Zoller},\ and\ \citenamefont
  {Phillips}}]{Cirac:1992cks}%
  \BibitemOpen
  \bibfield  {author} {\bibinfo {author} {\bibfnamefont {J.~I.}\ \bibnamefont
  {Cirac}}, \bibinfo {author} {\bibfnamefont {R.}~\bibnamefont {Blatt}},
  \bibinfo {author} {\bibfnamefont {P.}~\bibnamefont {Zoller}}, \ and\ \bibinfo
  {author} {\bibfnamefont {W.~D.}\ \bibnamefont {Phillips}},\ }\href@noop {}
  {\bibfield  {journal} {\bibinfo  {journal} {Physical Review A}\ }\textbf
  {\bibinfo {volume} {46}},\ \bibinfo {pages} {2668} (\bibinfo {year}
  {1992})}\BibitemShut {NoStop}%
\bibitem [{\citenamefont {deLaubenfels}\ \emph {et~al.}(2015)\citenamefont
  {deLaubenfels}, \citenamefont {Burkhardt}, \citenamefont {Vittorini},
  \citenamefont {Merrill}, \citenamefont {Brown},\ and\ \citenamefont
  {Amini}}]{deLaubenfels:2015fbs}%
  \BibitemOpen
  \bibfield  {author} {\bibinfo {author} {\bibfnamefont {T.~E.}\ \bibnamefont
  {deLaubenfels}}, \bibinfo {author} {\bibfnamefont {K.~A.}\ \bibnamefont
  {Burkhardt}}, \bibinfo {author} {\bibfnamefont {G.}~\bibnamefont
  {Vittorini}}, \bibinfo {author} {\bibfnamefont {J.~T.}\ \bibnamefont
  {Merrill}}, \bibinfo {author} {\bibfnamefont {K.~R.}\ \bibnamefont {Brown}},
  \ and\ \bibinfo {author} {\bibfnamefont {J.~M.}\ \bibnamefont {Amini}},\
  }\href@noop {} {\bibfield  {journal} {\bibinfo  {journal} {Physical Review
  A}\ }\textbf {\bibinfo {volume} {92}},\ \bibinfo {pages} {061402} (\bibinfo
  {year} {2015})}\BibitemShut {NoStop}%
\end{thebibliography}
\end{document}